\documentclass[letterpaper,11pt]{article}
\pdfoutput=1
\usepackage{jheppub}
\usepackage{cancel}
\usepackage{graphicx}
\usepackage{subfigure}
\usepackage{soul}
\usepackage{amsmath}
\usepackage{multirow}
\usepackage{hyperref}
\usepackage{epstopdf}
\usepackage{verbatim}
\def\lsim{\mathrel{\raise.3ex\hbox{$<$\kern-.75em\lower1ex\hbox{$\sim$}}}}
\def\gsim{\mathrel{\raise.3ex\hbox{$>$\kern-.75em\lower1ex\hbox{$\sim$}}}}

\def\beq{\begin{equation}}
\def\eeq{\end{equation}}
\def\be{\begin{equation}}
\def\ee{\end{equation}}
\def\bea{\begin{eqnarray}}
\def\eea{\end{eqnarray}}

\def\etmiss{\cancel{E}_{T}}

\title{Higgs Boson Decay to Light Jets at the LHC}
\author[a]{Linda M. Carpenter,}
\author[b,c,d]{Tao Han,}
\author[a]{Khalida Hendricks,}
\author[b]{Zhuoni Qian,}
\author[c, d]{Ning Zhou}
\date{\today}

\affiliation[a]{Department of Physics, Ohio State University
191 West Woodruff Ave., Columbus OH, 43210, U.S.A.}
\affiliation[b]{Pittsburgh Particle Physics, Astrophysics, and Cosmology Center,
Department of Physics and\\
Astronomy, University of Pittsburgh,
3941 O'Hara St., Pittsburgh, PA 15260, U.S.A.}
\affiliation[c]{Department of Physics, Tsinghua University, P.R.~China}
\affiliation[d]{Collaborative Innovation Center of Quantum Matter, Beijing, China}

\emailAdd{lmc@physics.osu.edu}
\emailAdd{than@pitt.edu}
\emailAdd{hendricks.189@osu.edu}
\emailAdd{zhq8@pitt.edu}
\emailAdd{nzhou2015@tsinghua.edu.cn}

\abstract{We study the Higgs boson $(h)$ decay to two light jets at the 14 TeV High-Luminosity-LHC (HL-LHC), where a light jet ($j$) represents any non-flavor tagged jet from the observational point of view.
The decay mode $h\to gg$ is chosen as the benchmark since it is the dominant channel in the Standard Model (SM), but the bound obtained is also applicable to the light quarks $(j=u,d,s)$. We estimate the achievable bounds on the decay branching fractions through the associated production $Vh\ (V=W^\pm,Z)$. Events of the Higgs boson decaying into heavy (tagged) or light (un-tagged) jets are correlatively analyzed.
We find that with 3000 fb$^{-1}$ data at the HL-LHC, we should expect approximately $1\sigma$ statistical significance on the SM $Vh(gg)$ signal in this channel. This corresponds to a reachable upper bound ${\rm BR}(h\to jj) \leq 4~ {\rm BR}^{SM}(h\to gg)$ at $95\%$ confidence level.
A consistency fit also leads to an upper bound
${\rm BR}(h\to cc) < 15~ {\rm BR}^{SM}(h\to cc)$ at $95\%$ confidence level.
The estimated bound may be further strengthened by
adopting multiple variable analyses, or adding other production channels. }


\preprint{
\begin{flushright}
PITT-PACC 1614 \\
\end{flushright}
}

\begin{document}
\maketitle
\flushbottom


\section{Introduction}

As we know for the Higgs detection at the LHC, $\gamma \gamma$ and $ZZ$ were the discovery channels for the Standard Model (SM)-like Higgs boson $(h)$ \cite{Aad:2012tfa,Chatrchyan:2012xdj}. Next came the $WW$ decay channel, all have been measured with more than $5\sigma$ significance at Run I by both experiments ATLAS \cite{ATLAS:2014aga} and CMS \cite{Chatrchyan:2013iaa}. While the $ZZ,WW$ channels are tree-level processes, most directly related to the electroweak symmetry breaking (EWSB) with the coupling strength proportional to $M_{W,Z}\sim gv$, the Higgs coupling to the top quark is best inferred from its contribution to the production $gg\to h$ and the decay $h\to \gamma\gamma$ with a fitted accuracy of around $30\%$ \cite{ATLAS:2016aa}. A direct measurement from Higgs and top associated production is yet to be established \cite{Khachatryan:2015ila, Aad:2016zqi}. For the lepton side, the challenging decay channel $h\to \tau^+\tau^-$ has also reached $5\sigma$ observation with a combined analysis of the two experiments \cite{ATLAS:2016aa}. With the upgrade of LHC to its higher center of mass energy at Run II and more accumulated data, the difficult mode $h \to b\bar b$ is expected to reach $5\sigma$ soon after several hundreds ${\rm fb}^{-1}$ at $14$ TeV \cite{atlasstudy}. Thus, the Higgs couplings to the heaviest generation of fermions will soon be settled to the values expected from the Standard Model (SM) prediction at an accuracy of about $20\%$ \cite{ATL-PHYS-PUB-2014-016}, and verifying the pattern of non-universal Yukawa couplings.

We next consider the LHC upgrade to a total integrated luminosity of 3000 fb$^{-1}$ at 14 TeV (HL-LHC).
While the precision measurements of those couplings will continue in the LHC experiments, it is imperative to seek  other ``rare decay'' channels, in the hope of uncovering any deviations from the SM. Among the rare channels, it is perhaps most promising to observe the clean mode $gg\to h \to \mu^+\mu^-$ \cite{Han:2002gp}, despite the small decay branching fraction BR$(h\to \mu^+\mu^-)\sim 2\times 10^{-4}$. A $5\sigma$ observation may be conceivable at the end of the run for HL-LHC with 3000 ${\rm fb}^{-1}$ \cite{ATL-PHYS-PUB-2014-016}, which would be of significant importance to establish the pattern of the Yukawa couplings by including a second generation fermion. For the other hadronic channels, it would be extremely challenging to make any measurements at the LHC due to the overwhelmingly large QCD backgrounds.\footnote{Due to the much cleaner experimental environment, a lepton collider such as International Linear Collider 
(ILC) \cite{Asner:2013qy} or a circular $e^+ e^-$ collider \cite{Koratzinos:2014cla, CEPC-SPPCStudyGroup:2015csa}, running at the $Zh$ threshold or higher energies, will give us much better sensitivity to the hadronic decays of the Higgs. The expected accuracy on $h\to gg$ and $h\to c\overline{c}$ will be $7\%\ (2.3\%)$ and $8.3\%\ (3.1\%)$ respectively, with the 250 GeV (1TeV) mission \cite{Asner:2013qy}.}

The most promising production mechanism for the hadronic decay signal of the Higgs boson is
\bea
\label{eq:vh}
pp \to V h,\ \  {\rm where} \ \ V=W^\pm,\ Z.
\eea
With $W/Z$ decaying leptonically to serve as effective triggers, the Higgs signal may be detected from the construction of its invariant mass of the hadronic products. To sufficiently suppress the large QCD backgrounds, it was proposed \cite{Butterworth:2008sd} to look for highly-boosted events for $h \to b\bar b$ against the leptonic $W/Z$.
Studies on these processes at HL-LHC shows a $\approx 20 \sigma\ (9 \sigma)$ significance for the signal $Vh, h\to b\bar b$, with statistical (systematic added) uncertainty estimated \cite{atlasstudy}.
%
%
Marching to the channel involving the second generation quarks, the sensitivity to $Vh, h\to c\bar c$ is  significantly worse. Bounds are extrapolated in a recast study in Ref.~\cite{Perez:2015lra} to be $\sim 6.5$ times the SM value (statistic errors assumed only). This is expected, given that BR$(h\to b\bar b)$ is $\sim 20$ times larger than BR$(h\to c\bar c)$, that expected $b$-tagging is twice as efficient as $c$-tagging, and that the dominant background $Vbb(cc)$ in the relevant kinematic region is about the same order.
An interesting proposal to search for $h\to J/\psi + \gamma$ \cite{Bodwin:2013gca} does not seem to increase the observability for $hcc$ coupling due to too low an event rate \cite{Aad:2015sda, ATL-PHYS-PUB-2015-043}.

It is natural to ask to what extent one would be able to search for other hadronic decays of the Higgs boson. We here quote the updated calculations of the branching fractions for the 125 GeV Higgs boson decay hadronically in the SM \cite{deFlorian:2016spz}
\bea
&& {\rm BR}(h\to b\bar b)=58.2\%,\quad {\rm BR}(h \to  c\bar c) = 2.89\%, \\
&& {\rm BR}(h\to gg)=8.18\%,\quad {\rm BR}(h\to u\bar u, d\bar d,s\bar s)< 0.03\% .
\eea
While the decay rates to light quarks predicted in the SM would be too small to be observable, the decay to a pair of gluons, mediated via the heavy top quark, will be nearly three times larger than the $c\bar c$ channel. The experimental signatures for those channels would be to search for the un-tagged light jet pairs $jj$, which form a mass peak near the Higgs boson mass $m_h$. Obviously, the lack of a heavy-flavor tag makes background suppression difficult. However, we point out that  the event sample so defined naturally exists and falls in to a class of mis-tagged events for $h\to b\bar b, c\bar c$ searches as well, that must be properly quantified with respect to the mis-tag rates as the ``contamination'' to the genuine decays of the Higgs boson to light jets.

%
%
%

In this work we set out to study Higgs decay to a pair of light un-tagged jets $h\to jj$, in the associated production channel as in Eq.~(\ref{eq:vh}).
We will exploit the leptonic final state decays of the electroweak gauge bosons,
and employ a hadronic tag for the Higgs boson while optimizing the mass reconstruction.
Evaluating the major sources of statistic (or systematic) uncertainties, we argue that a $1\sigma$ sensitivity of 1 (or 4) times the SM value can be achieved for the case where the Higgs decays to un-tagged jets.  This is achieved with a judicious choice of kinematic discriminants and a combination of the final state channels.
Together with $h\to b\bar b$ and $h\to c\bar c$ studies, the un-tagged channel puts an independent dimension of bound in the space of branching ratios of Higgs decays to quarks and gluons. Assuming a well measured $ggh$ coupling at the end of HL-LHC \cite{ATL-PHYS-PUB-2014-016}, the result further puts comparable but independent constraints on the light-quark Yukawa couplings. We also estimate that this channel may offer a better probe to the strange-quark Yukawa coupling.




This paper proceeds as follows, Section \ref{sec:sigbkg} specifies the signal and dominant background processes. Section \ref{sec:selection} describes and presents the detailed analyses  and gives the main results in terms of the cut-efficiency tables and figures.  In the same section, we also study how to control the systematic errors for the large backgrounds. Section \ref{sec:alternatives} describes an alternate search strategy based on momentum balance discriminants.  Section \ref{sec:results} calculates the signal sensitivity and presents obtained constraints on Higgs couplings to quarks and gluons in a correlated manner, while Section~\ref{sec:conclude} summarizes and concludes.



\section{Signal and Background Processes}
\label{sec:sigbkg}



\begin{figure}[t]
	\centering
	       \includegraphics[scale=0.5]{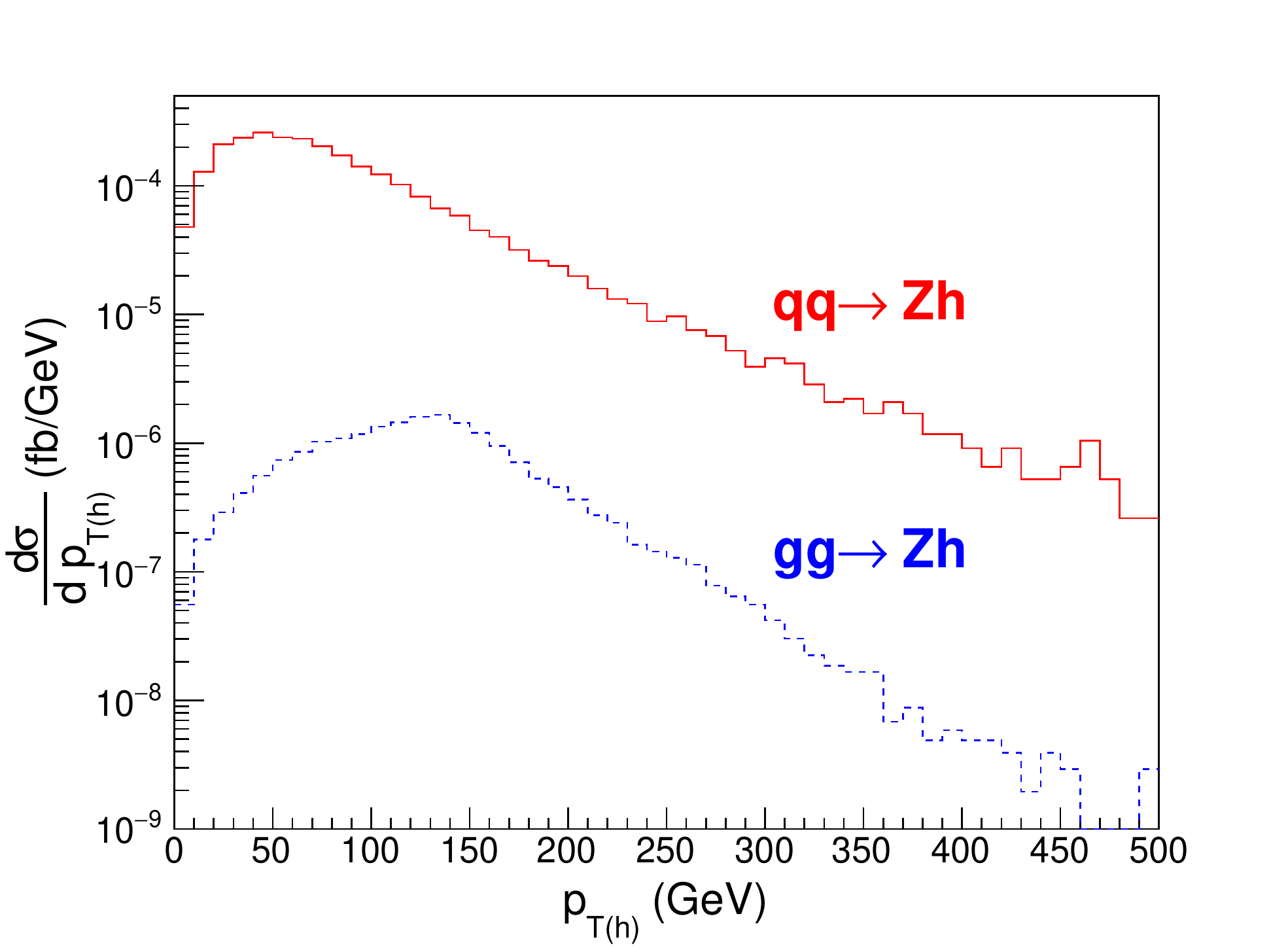}
	\caption[]{Higgs boson transverse momentum distribution for the signal processes $qq\to Zh$ (upper solid curve) and $gg\to Zh$ (lower dashed curve) at the 14 TeV LHC.
	}
	\label{fig:pt-higgs-log}
\end{figure}

As discussed above, the promising channel in which to study the Higgs decay to light jets is the associated production with an electroweak gauge boson $W$ or $Z$, which subsequently decays to leptons. Depending on the production mechanisms and the final states, we consider the following subprocesses
%
\bea
\label{eq:w}
&&q\bar q \to W^\pm h \to \ell^\pm \nu + jj , \\
\label{eq:z}
&&q\bar q,\ gg \to  Zh\to \left\{
\begin{array}{ll}
 \ell^+\ell^- + jj , & \\
\nu\bar \nu+ jj,  &
\end{array}
\right.
\eea
where $\ell=e,\mu$ and $j=g$ or $u,d,s$. Practically, $j$ is a gluon as expected in the SM. We thus generically denote the SM signal by $Vh(gg)$, whenever convenient.
In our calculations, events are generated with MadGraph at the leading order, with ``NN23NLO" as the PDF set.
For the $gg\to Zh$ process via the quark loops, we use Madgraph\_NLO \cite{Alwall:2014hca} and Madspin \cite{Artoisenet:2012st}.
This channel contributes about $10\%-20\%$ to the total $Zh$ production rate.
We apply an overall rescaling of QCD K-factors to the signal processes, to match the total NNLO QCD and NLO EW cross section results taken from summary of Higgs cross section working 
group \cite{deFlorian:2016spz}. The K-factors are about 2 and 1.2 for the $gg$ and $q\bar q$, respectively. We have included the finite masses for the fermions running in the loop in the $gg$ initiated process.
Some care is needed regarding the $gg$ process because of its different transverse momentum ($p_T$) dependence and sensitivity to new physics contribution in the loop as discussed 
in Ref.~\cite{Englert:2013yo}.
In Fig.~\ref{fig:pt-higgs-log}, we compare the Higgs boson transverse momentum distributions for the signal processes $q\bar q \to Zh$ and $gg\to Zh$.
The $q\bar q$-initiated channel peaks at $p_{T(h)}\approx 50\ {\rm GeV}$, a typical mass scale associated with the final state particles of $Zh$.
The $gg$-initiated channel peaks at around $p_{T(h)} \approx 150\ {\rm GeV}$, due to the top mass threshold enhancement. The differential cross section of $gg$ drops faster than $q\bar q$ with increasing $p_{T(h)}$, due to the destructive interference between the triangle and box diagrams.

The Higgs is further decayed according to the branching ratios listed in Ref.~\cite{deFlorian:2016spz}. Events are then showered and hadronized using PYTHIA6 \cite{Sjostrand:2006za}, and run through DELPHES \cite{deFavereau:2013fsa} for detector simulation and jet reconstruction.
For the SM backgrounds, we mainly consider the dominant irreducible background process $V+jj$ at LO, where the $V$ decays and contributes accordingly to the three signal channels.
At the generator level, we apply some basic cuts on the jets to remove infrared and collinear divergences for the QCD background processes
\be
p_{T(j)}>20\ {\rm GeV}, \quad |\eta_j| < 3, \quad R_{jj} > 0.4.
\label{eq:cutjl}
\ee
The hadronic jets are reconstructed with anti-kt jet algorithm with a cone size $R=0.4$.
In our future analyses, we will be considering a relatively boosted Higgs recoiling off of the vector boson.  Therefore, to improve the simulation statistics, we also add a generator-level cut on the vector boson
\be
p_{T(V)}>150\ {\rm GeV}.
\label{eq:cutV}
\ee

In Table \ref{tab:parton-xsec} we give the cross sections used for our signal and background processes including the basic cuts in Eq.~(\ref{eq:cutjl}) and
with various $p_T$ thresholds for the vector boson. The first is the total cross section with no $p_{T(V)}$ cut, the second and third demand $p_{T(V)}$ cuts of 150 and 200 GeV respectively. No cuts on the final state leptons are applied for the table.

\begin{table}
\centering
\begin{tabular}{|c|c|c|c|}
\hline
$\sigma$ (fb) & cuts Eq.~(\ref{eq:cutjl})  & $+$ Eq.~(\ref{eq:cutV}) & $+\ p_{T(V)}>200$ GeV \\ \hline
$q\bar q\to Z h \to \ell^{+}\ell^-\ gg$ & 3.5 & 0.39 & 0.17 \\ \hline
$gg\to Z h \to \ell^+\ell^-\ gg$ & 0.71 & 0.20 & $6.2\times 10^{-2}$ \\ \hline
$q\bar q\to Z j j \to \ell^+\ell^-\ jj$ & $2.5\times 10^5$ & $1.2\times 10^4$ & $4.8\times 10^3$ \\ \hline \hline
$q\bar q\to W h \to \ell\nu\ gg$ & 20 & 2.3 & 0.99 \\ \hline
$q\bar q\to W j j \to \ell\nu\ jj$ & $2.5\times 10^6$ & $1.0\times 10^5$ & $3.9\times 10^4$ \\ \hline
$pp\to t\bar t\to \ell\nu jjb\bar b$ & $1.1\times 10^5$ & $1.5\times 10^4$ & $5.7\times 10^3$ \\ \hline  \hline
$q\bar q\to Z h \to \nu\nu\ gg$ & 11 & 1.2 & 0.50 \\ \hline
$gg\to Z h \to \nu\nu\ gg$ & 2.1 & 0.60 & 0.18 \\ \hline
$q\bar q\to Z jj \to \nu\nu\ jj $ & $7.4\times 10^5$ & $3.6\times 10^4$ & $1.4\times 10^4$\\ \hline
\end{tabular}
\caption[]{Cross sections in units of fb for signal and dominant background processes, with the parton-level cuts of Eq.~(\ref{eq:cutjl}), and boosted regions $p_{T(V)}>150,\ 200$ GeV.}
\label{tab:parton-xsec}
\end{table}
%
\section{Signal Selection}
\label{sec:selection}

\begin{figure}[tb]
	\centering
	       \includegraphics[scale=0.7]{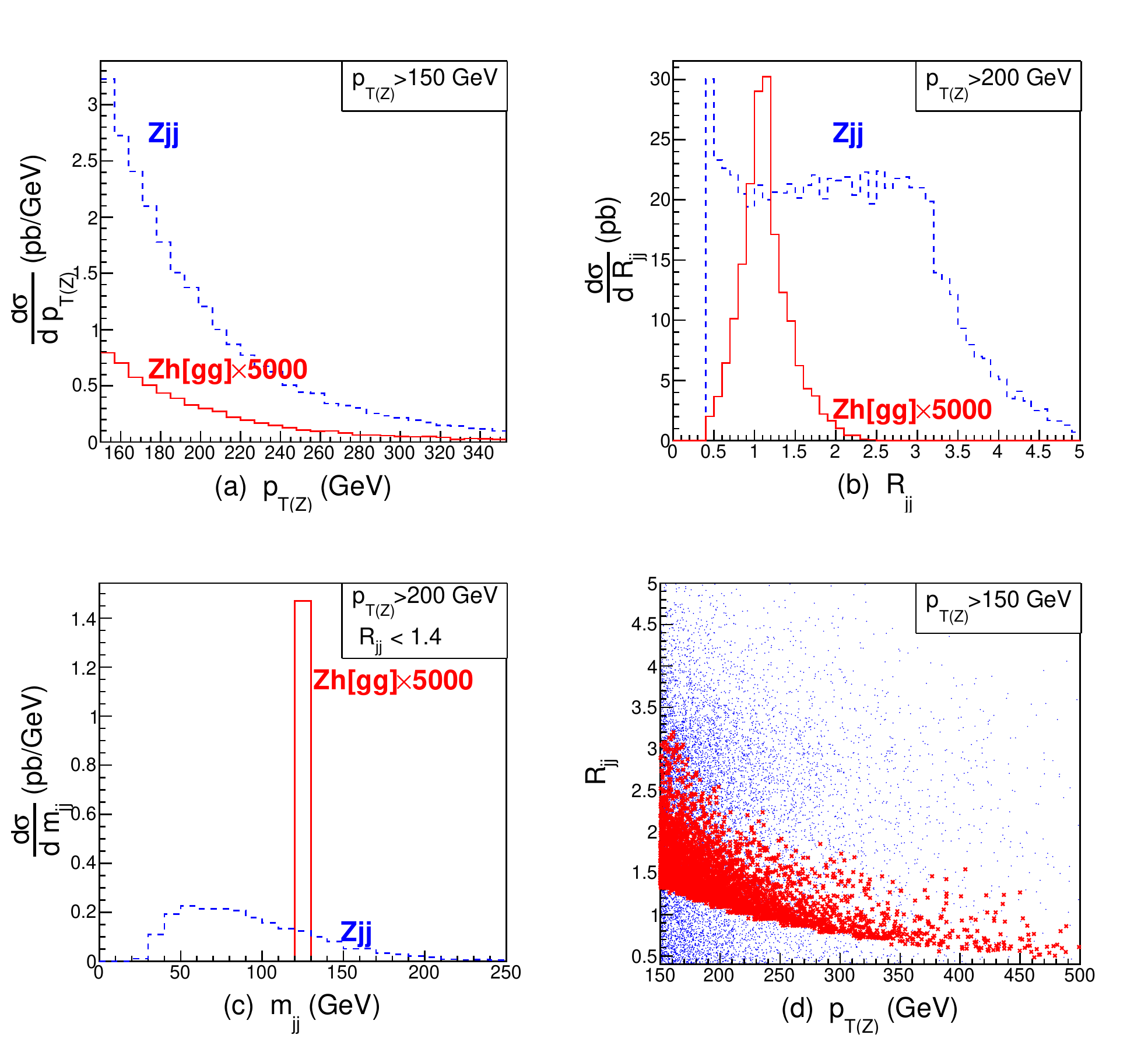}
	\caption[]{Kinematical distributions of the signal process $pp\to Zh, h\to gg$ (solid curves, scaled up by a factor of 5000) and the leading background $pp\to Zjj$ (dashed curves) for (a) $p_{T(Z)}$, (b) $R_{jj}$, (c) $m_{jj}$, and (d) event scatter plot in $R_{jj}-p_{T(Z)}$ plane, with the (red) dense band with crosses as the signal events and (blue) dots as the background. Generator level cuts of Eqs.~(\ref{eq:cutjl}) and (\ref{eq:cutV}) have been applied.
	}
	\label{fig:pt-rgg}
\end{figure}

In further studying the signal characteristics in Eqs.~(\ref{eq:w}) and (\ref{eq:z}), we categorize the channels according to the zero, one, or two charged leptons from the vector boson decays. In addition, the signal has two leading jets from the Higgs decay, with invariant mass of the Higgs boson.  At high $p_{T(h)}$, the distance between the two hadronic jets can be estimated as
\be
R_{jj} \approx \frac{1}{\sqrt{z(1-z)}}\frac{m_h}{p_{T(h)}},
\label{eq:R}
\ee
where $z,~ 1-z$ are the momentum fraction of the two jets.
The LO parton-level distributions of three kinematic discriminants for the $Zh$ channel, the transverse momentum $p_{T(Z)}$, the jet separation $R_{jj}$, and the di-jet invariant mass $m_{jj}$, are shown in Fig.~\ref{fig:pt-rgg}, comparing the signal (solid) and dominant background (dashed), after the generator-level  cuts as in Eqs.~(\ref{eq:cutjl}) and (\ref{eq:cutV}). Obviously, $p_{T(Z)}$ is singular for the QCD background as seen in Fig.~\ref{fig:pt-rgg}(a). The two jet separation $R_{jj}$ in Fig.~\ref{fig:pt-rgg}(b) shows the either collinear feature from the parton splitting in the final state radiation (FSR) or back-to-back near $\pi$ due to the initial state radiation (ISR) for the background process, and is narrowly populated near $2m_h/p_{T(h)}$ for the signal. The resonance bump near $m_h$ is evident as in Fig.~\ref{fig:pt-rgg}(c). Because of the small rate, the signal curves have been scaled up by a factor of 5000.
We also show an event scatter plot in Fig.~\ref{fig:pt-rgg}(d), where the (red) dense band with crosses presents the signal events and the (blue) dots show the background events. We see the strong correlation between the boosted $p_{T(Z)}$ and collimated jets with smaller $R_{jj}$.

To suppress the huge QCD di-jet backgrounds, we must optimize the reconstruction of the Higgs mass.
There are two common methods to reconstruct hadronic decays of Higgs boson depending on the kinematical configurations. One is the sub-structure (fat-jet) approach: an early example for Higgs search in $b\bar b$ channel was introduced in Ref.~\cite{Butterworth:2008sd}. Because of the highly boosted nature of the Higgs boson, a fat-jet identified as the hadronic decay products of the Higgs boson is first selected. Various jet substructure observables and techniques such as mass-drop and filtering \cite{Butterworth:2008sd}, pruning \cite{Ellis:2009su}, trimming \cite{Krohn:2009th}, N-subjettiness \cite{Thaler:2010tr} etc. can be applied on the fat-jet to further improve the reconstruction of the invariant mass.
The other approach is to simply resolve the leading jets. This is the common practice when the Higgs is produced not far from the threshold, and the Higgs is identified as the sum of the two leading jets.
Experimentally, the anti-kt jet algorithm, given its regular jet shape, gives good reconstruction of hadronic jets, and is the default hadronic jet reconstruction algorithm used at ATLAS/CMS. The $Vh(b\bar b)$ search at LHC is currently carried with the two resolved jet with anti-kt $R=0.4$ method.
In a recent analysis \cite{Butterworth:2015bya} the two methods are compared for the $Wh, h\to bb$ process for LHC14 in the kinematic region $200\ {\rm GeV} < p_{T(h)} < 600\ {\rm GeV}$. The resolved approach is better in the $200\ {\rm GeV} < p_T < 300 \ {\rm GeV}$ range. The jet-substructure approach is significantly better in the $p_T > 600\  {\rm GeV}$. The results are  qualitatively expected, since the high $p_T$ corresponds to a smaller cone-size of the fat-jet as argued in Eq.~(\ref{eq:R}).

\begin{figure}[t]
	\centering
	       \includegraphics[scale=0.8]{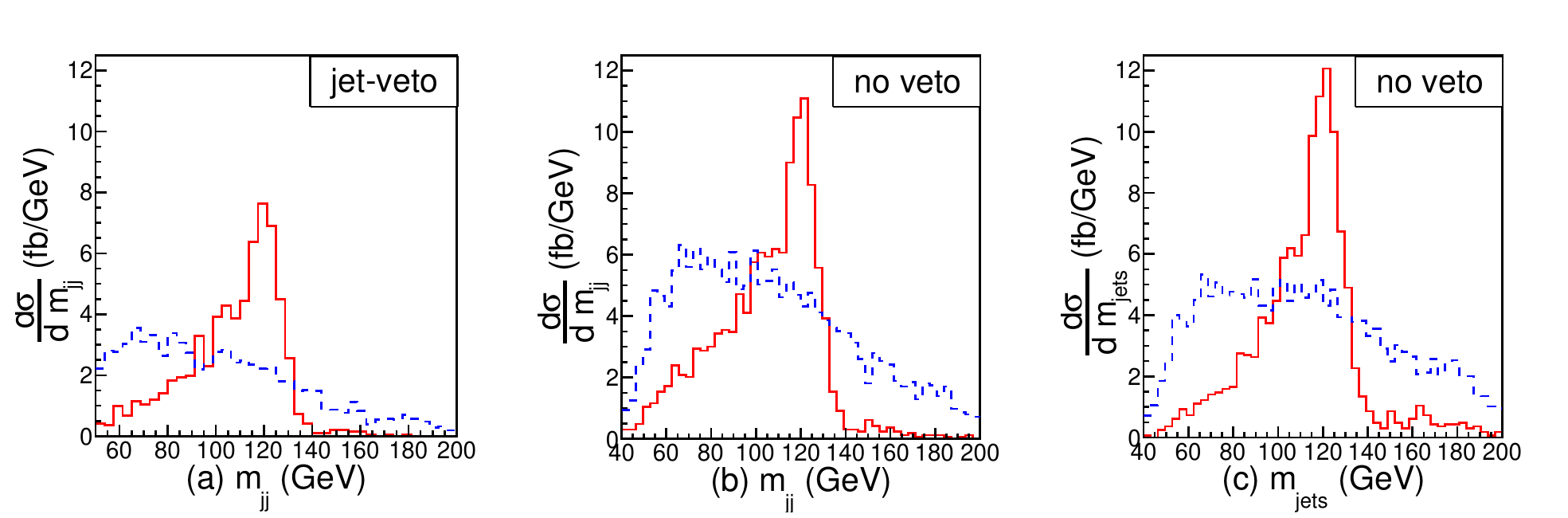}
	\caption[]{Invariant mass distributions $m_{jj}$ of the signal process $pp\to Zh, h\to gg, Z\to\ell\ell$ (solid curves, scaled up by a factor of 5000) and the leading background $pp\to Zjj$ (dashed curves) for (a) with 2 jets only, (b) with 2 leading jets to reconstruct $m_{jj}$, (c) with 2 leading jets plus other jets together to reconstruct $m_{jets}$. All selection cuts as in Sec.~\ref{item:2l-cuts} except for ${\rm m}_h$ cut are applied.
	}
	\label{fig:hmass-reconst}
\end{figure}

Since the signal events tend to populate near the kinematic threshold, we will exploit the resolved method with two hard jets.
However, additional QCD radiations from the highly energetic jets are not negligible. Kinematically, it gives a reconstructed di-jet mass peak smeared towards lower value. Some related effects including the NLO correction is studied in Ref.~\cite{Banfi:2012kq}. We thus propose a modification of the two-jet-resolved method by including possible additional jets in the decay neighborhood -- a ``resolved  Higgs-vicinity" method. After clustering the jets with anti-kt $\Delta R=0.4$,
two leading $p_T$ jets are clustered as the ``Higgs-candidate". Then additional jets $j'$ are also clustered to the ``Higgs candidate" in sequence of angular vicinity, whenever $R_{Hj'} \leq R_{\rm max}$. For the rest of the analyses, we choose
\be
R_{\rm max} =1.4.
\label{eq:Rmax}
\ee
The optimal method is to select events with two leading $p_T$ jets that satisfy $R_{jj} \leq R_{\rm max}$, and add to the di-jet system any sub-leading jets within the distance $R_{\rm max}$. In practice, we find that including one additional hard radiation in the decay is sufficient.
In Fig.~\ref{fig:hmass-reconst} we compare several resolved-jet methods in their reconstruction of the Higgs mass, against the $Vjj$ background. The central and hard jet requirements are $p_{T(j)}> 30$ GeV and $|\eta_{j}|<2.5$. In Fig.~\ref{fig:hmass-reconst}(a), we reconstruct the Higgs with the two leading $p_T$ jets and veto events with more than two central hard jets. As shown in the plot, the veto method removes the background most efficiently, the cut also reduces the signal significantly. Fig.~\ref{fig:hmass-reconst}(b) shows the 2jet-inclusive case, which is the same as (a) but does not veto additional jets. It improves the signal rate, but the signal mass peak is still smeared to the lower value. Fig.~\ref{fig:hmass-reconst}(c) is the ``resolved  Higgs-vicinity" method, which adds the additional hard jet, and sharpens the mass peak to help increase the overall $S/\sqrt{B}$ sensitivity.

%

\begin{figure}[t]
	\centering
	  \includegraphics[width=0.45\linewidth]{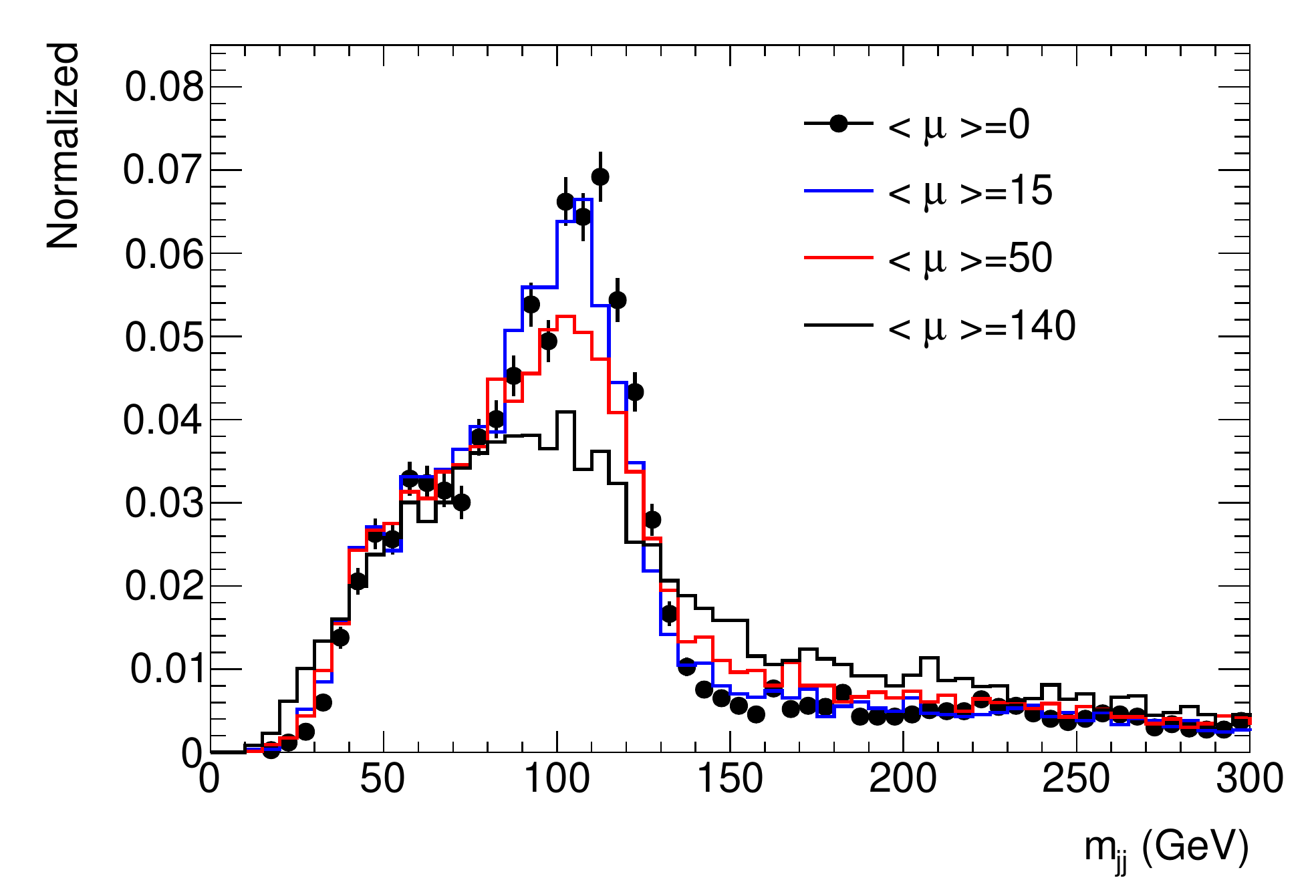}
  	\includegraphics[width=0.45\linewidth]{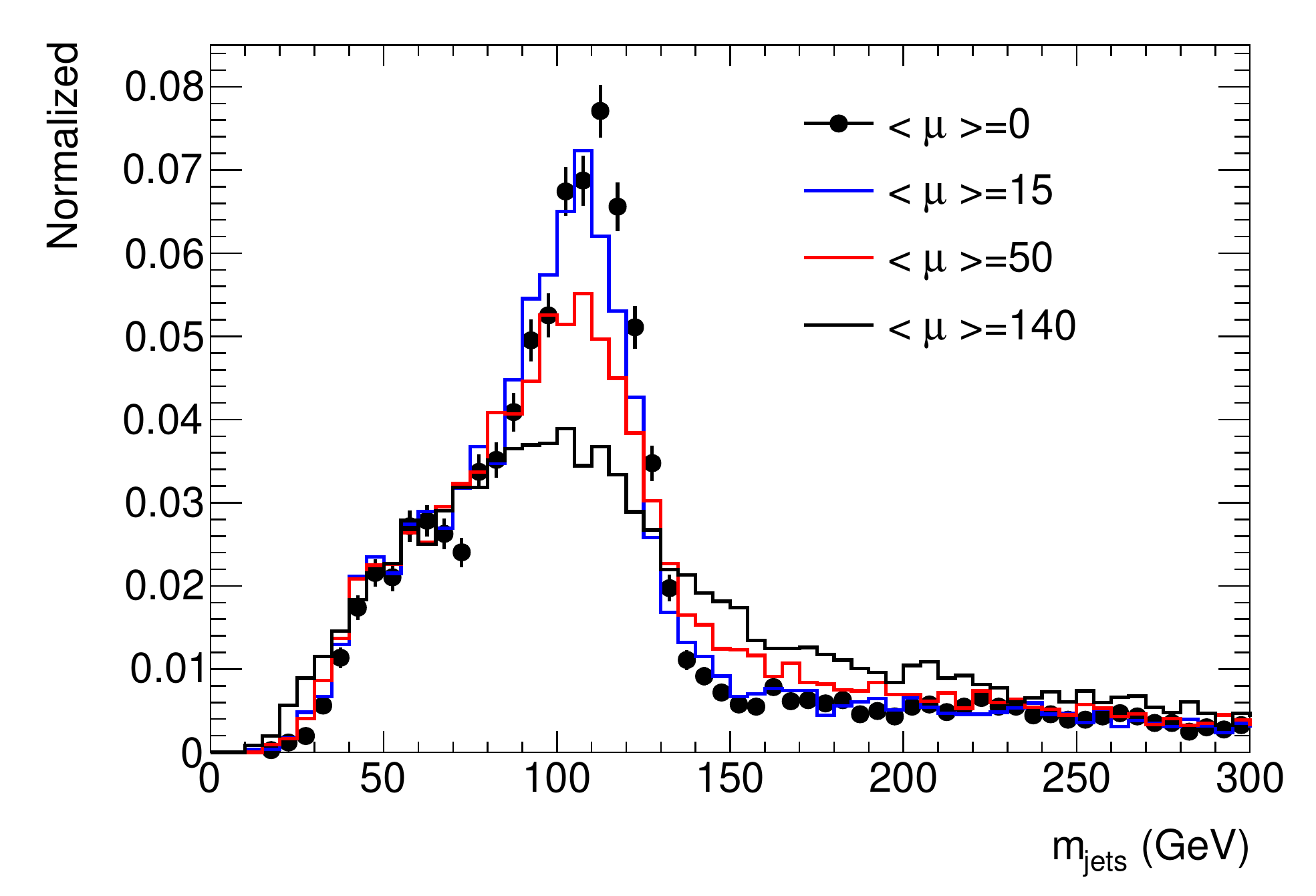}
	\caption[]{Invariant mass distributions constructed from (a) two-jet events and (b) three-jet events with different pile-up values $\langle \mu \rangle = 0, 15, 50, 140$, respectively.
	}
	\label{fig:pile-up}
\end{figure}

%
We study the sensitivity to pile-up contamination of this reconstruction method. In Fig.~\ref{fig:pile-up}, we compare it with the two jet resolved method adding pile-up samples in DELPHES. As expected, the additional-jet method is more sensitive to the pile-up jets, yet still retains a slight advantage even under pile-up value $\langle \mu \rangle =140$ \cite{ATL-UPGRADE-PUB-2013-014}.

%
%

In the following, we describe the searches with the detailed signal and background analyses, for the channels with two, one and zero charged leptons, respectively. For simplicity, we use 2 jets reconstruction of the mass peak from now on.

\subsection{$\ell^+ \ell^- + jj$ channel}
\label{sec:2lepton}

For the two-lepton channel, we simulate the signal processes as in Eq.~(\ref{eq:z}) with $Z\to \ell^+\ell^-, \ h\to gg$. We require exactly one pair of charged leptons $\ell^\pm=e^{\pm}$ or $\mu^{\pm}$, same flavor, opposite charge, along with at least two energetic jets.
The dominant background is by far from $Z+jj$.
The two leading $p_T$ jets are required to be close by having a separation less than $R_{\rm max}= 1.4$, and an invariant mass between 95 and 150 GeV. They satisfy the following acceptance cuts
\begin{itemize}
\item{2 leptons with $p_{T(l)}> 30 $ GeV and $|\eta_{l}|<2.5$ }
\item {$p^{}_{T(\ell\ell)}> 200$ GeV}
\item{at least 2 jets  with $p_{T(j)}> 30$ GeV and $|\eta_{j}|<2.5$ }
\item {$R_{j_1j_2}<1.4$}
\item{$95$ GeV$<m_{h}<150$ GeV}
\label{item:2l-cuts}
\end{itemize}
%
The di-jet mass window around $m_h$ is chosen to optimize the $S/\sqrt B$ at HL-LHC.
Table \ref{tab:lljj-cutflow} shows the efficiency of applying the sequence of cuts. The overall efficiencies are about $14\% ,\ 7.6\%$, for the $q\bar q,\ gg$ initiated signal processes, respectively, and about $1.9\%$ for the background process. We would like to point out that from only the statistical sense, the signal sensitivity $S/\sqrt B$ would not be notably increased from the generator level results to that with final cuts. However, the fact that the background is reduced by around two orders of magnitude helps to control the systematic uncertainties, as we will discuss later.
%

\begin{table}[tb]
\centering
\begin{tabular}{|c|c|c|c|}
\hline
cut eff (\%) & $q\bar q\to Z h \to \ell^{+}\ell^- gg$ & $gg\to Z h \to \ell^+\ell^- gg$ & $q\bar q\to Z j j \to \ell^+\ell^- jj$ \\ \hline
$\sigma$ (fb) & $3.9\times 10^{-1}$ & $2.0\times 10^{-1}$ & 1.2$\times 10^4$ \\ \hline
2 leptons & 59\% & 52\% & 40\% \\ \hline
$\geq 2$~jets & 51\% & 49\% & 32\% \\ \hline
$70<m_{ll}<110$ & 50\% & 49\% & 31\% \\ \hline
$p_{T(\ell\ell)}>200$ GeV & 26\% & 23\% & 16\% \\ \hline
$R_{j_1j_2}<1.4$  & 21\% & 12\% & 5.3\% \\ \hline
$95<m_{h}<150$ GeV & 14\% & 7.6\% & 1.9\% \\ \hline
final (fb) & $5.4\times 10^{-2}$ & $1.5\times 10^{-2}$ & 2.4$\times 10^2$ \\ \hline
\end{tabular}
\caption[]{
The consecutive cut efficiencies for signal $\ell^+\ell^-\ jj$ and dominant background processes at the LHC.}
\label{tab:lljj-cutflow}
\end{table}
%

%

\subsection{$\ell^\pm + \etmiss + jj$ channel}
\label{sec:1lepton}
For the one-lepton channel, we look at signal process in Eq.~(\ref{eq:w}) with $W\to \nu\ell,\ h\to gg$. The dominant backgrounds are $W+jj$ and $t\bar t$. Similar to the last section, the acceptance  cuts are
\begin{itemize}
\item{one lepton $p_{T(\ell)}> 30$ GeV and $|\eta_{\ell}|<2.5$}
\item {$p^{}_{T(\nu\ell)}> 200$ GeV}, \ \  {$\etmiss > 30$ GeV }
\item{at least 2 jets with $p_{T(j)}> 30$ GeV and $|\eta_{j}|<2.5$}
\item{ $R_{j_1j_2}<1.4$}
\item {$95$ GeV $<m_{h}<150$  GeV}.
\end{itemize}
The $W$ transverse momentum $p_{T(\nu\ell)}$ can be reconstructed from the charged lepton plus the missing transverse momentum $\etmiss$.
Table \ref{tab:lvgg-cutflow} shows the cut-flow at various stages of the cuts applied.
The overall efficiencies are about $18\%$ for the $q\bar q$ initiated signal process,
and about $2.5\%,\ 2.5\%$ for the $Wjj,\ t\bar t$ background processes, respectively.
%
%
%
\begin{table}[tb]
\centering
\begin{tabular}{|c|c|c|c|}
\hline
cut eff (\%) &$q\bar q\to W h \to \ell\nu gg$ & $q\bar q\to W j j \to \ell\nu jj$ & $t\bar t \to \ell\nu jj b\bar b$ \\ \hline
$\sigma$ (fb) & 2.3 & 1.0$\times 10^5$ & 1.5$\times 10^4$ \\ \hline
$\etmiss>30$ GeV & 94\% & 87\% & 93\% \\ \hline
1 lepton & 72\% & 52\% & 62\% \\ \hline
$p_{T(\ell\nu)}>200$ GeV & 39\% & 24\% & 26\% \\ \hline
$\geq2$ jets & 35\% & 20\% & 22\% \\ \hline
$R_{j_1j_2}<1.4$ & 27\% & 6.8\% & 11\% \\ \hline
$95<m_{h}<150$ GeV  & 18\% & 2.5\% & 2.5\% \\ \hline
final (fb) & $4.1\times 10^{-1}$ & $2.5\times 10^3$ & $3.7\times 10^2$ \\ \hline
\end{tabular}
\caption[]{The consecutive cut efficiencies for signal $\ell^\pm \etmiss\ jj$ and dominant background processes at the LHC.}
\label{tab:lvgg-cutflow}
\end{table}
%
\subsection{$\etmiss + jj$ channel}
\label{sec:0lepton}

The zero-lepton channel is studied with signal processes as in Eq.~(\ref{eq:z}) with $Z\to \nu\nu,\ h\to gg$. The dominant background again mainly is $Z+jj$. Similar to the above, the cuts acceptance are
\begin{itemize}
\item{lepton veto with $p_{T(\ell)}> 30$ GeV $|\eta_{\ell}|<2.5$}
\item {$\etmiss  > 200$ GeV}
\item{at least 2 jets with $p_{T(j)}> 30$ GeV $|\eta_{j}|<2.5$}
\item {$R_{j_1j_2}<1.4$}
\item {$95$  GeV $<m_{h}<150 $  GeV}.
\end{itemize}
The $\etmiss$ is essentially from $p^{}_{T(Z)}$.
Table \ref{tab:vvjj-cutflow} shows the cut-flow at various stages of the cuts applied.
The overall efficiencies are about $23\% ,\ 15\%$, for the $q\bar q,\ gg$ initiated signal processes, respectively, and about $4.5\%$ for the background process.

Results presented in the above three sections have been double checked by other approaches.

\begin{table}[tb]
\centering
\begin{tabular}{|c|c|c|c|}
\hline
cut eff (\%) & $q\bar q\to Z h \to \nu\nu gg$ & $gg\to Z h \to \nu\nu gg$ & $q\bar q\to Z jj \to \nu\nu jj$\\ \hline
$\sigma$ (fb) & 1.2 & $6.0\times 10^{-1}$ & 3.6$\times 10^4$ \\ \hline
$\etmiss>200$ GeV & 49\% & 44\% & 42\% \\ \hline
$\geq2$ jets  & 45\% & 43\% & 35\% \\ \hline
$R_{j_1j_2}<1.4$ & 36\% & 25\% & 12\% \\ \hline
$95<m_{h}<150$ GeV  & 23\% & 15\% & 4.5\% \\ \hline
final (fb) & $2.7\times 10^{-1}$ & $8.9\times 10^{-2}$ & $1.6\times 10^3$ \\ \hline
\end{tabular}
\caption[]{The consecutive cut efficiencies for signal $\etmiss\ jj$ and dominant background processes at the LHC.}
\label{tab:vvjj-cutflow}
\end{table}
%

\subsection{Background control}
\label{sec:bkg-fit}

As calculated earlier and presented in the previous tables, the signals for $h\to gg$ in the SM associated with $W/Z$ to leptons at the 3000 fb$^{-1}$ HL-LHC may lead to sizable event rates, with about
200 events for the $\ell^+\ell^-$ channel,
1300 events for the $\ell^\pm\nu$ channel, and
1200 events for the $\nu\nu$ channel, respectively. However, the difficulty is the overwhelmingly large SM background, with a signal-to-background ratio at the order of $10^{-4}$. As such, one must be able to control the systematic errors to sub-percent in order to reach statistically meaningful result. This is an extremely challenging job, and one would not be able conclude without real data to show the detector performances. On the other hand, there are ideas to shoot at the goal. Here we adopt one of the commonly considered methods and demonstrate our expectations.

\begin{figure}[tb]
  \center
  \includegraphics[width=0.49\linewidth]{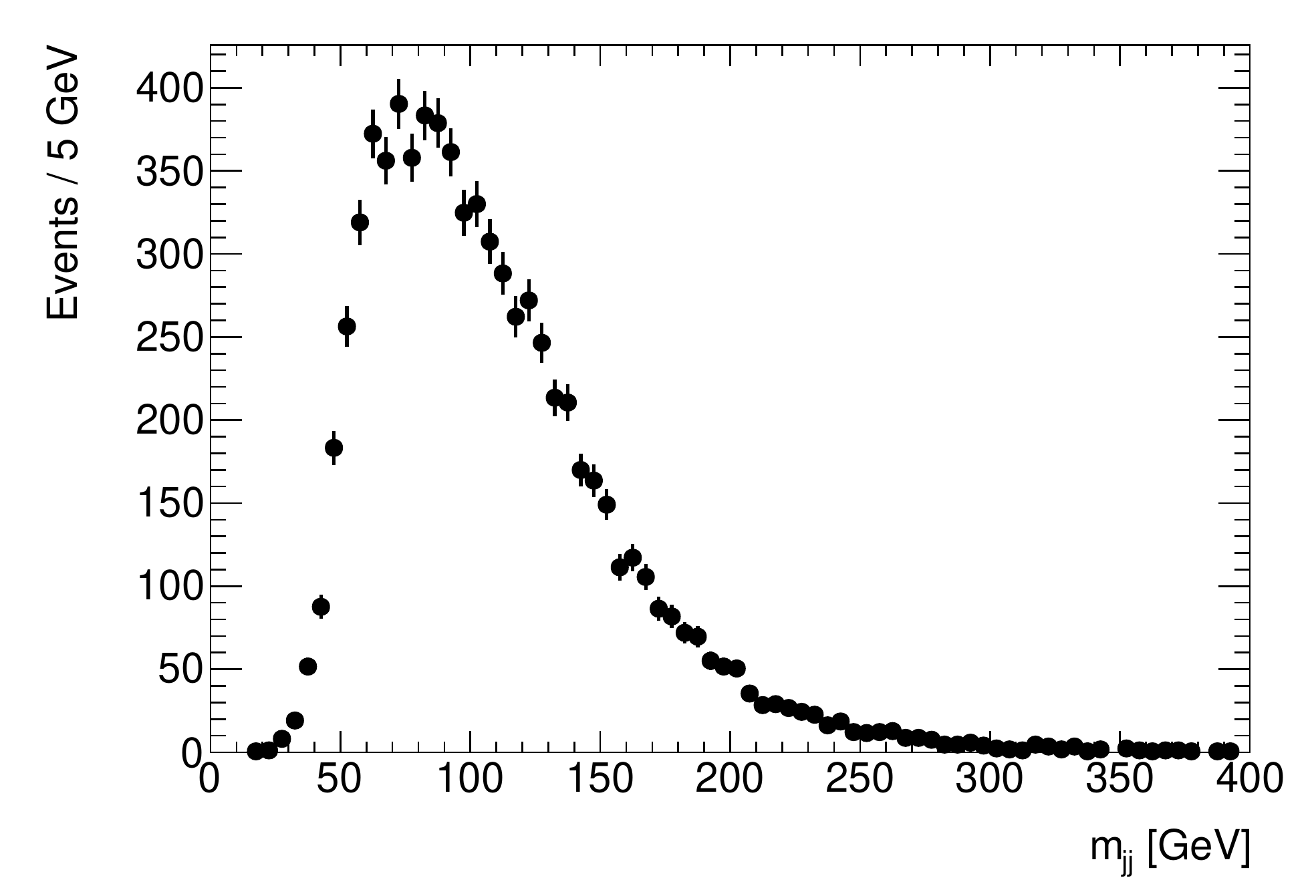}
  \includegraphics[width=0.49\linewidth]{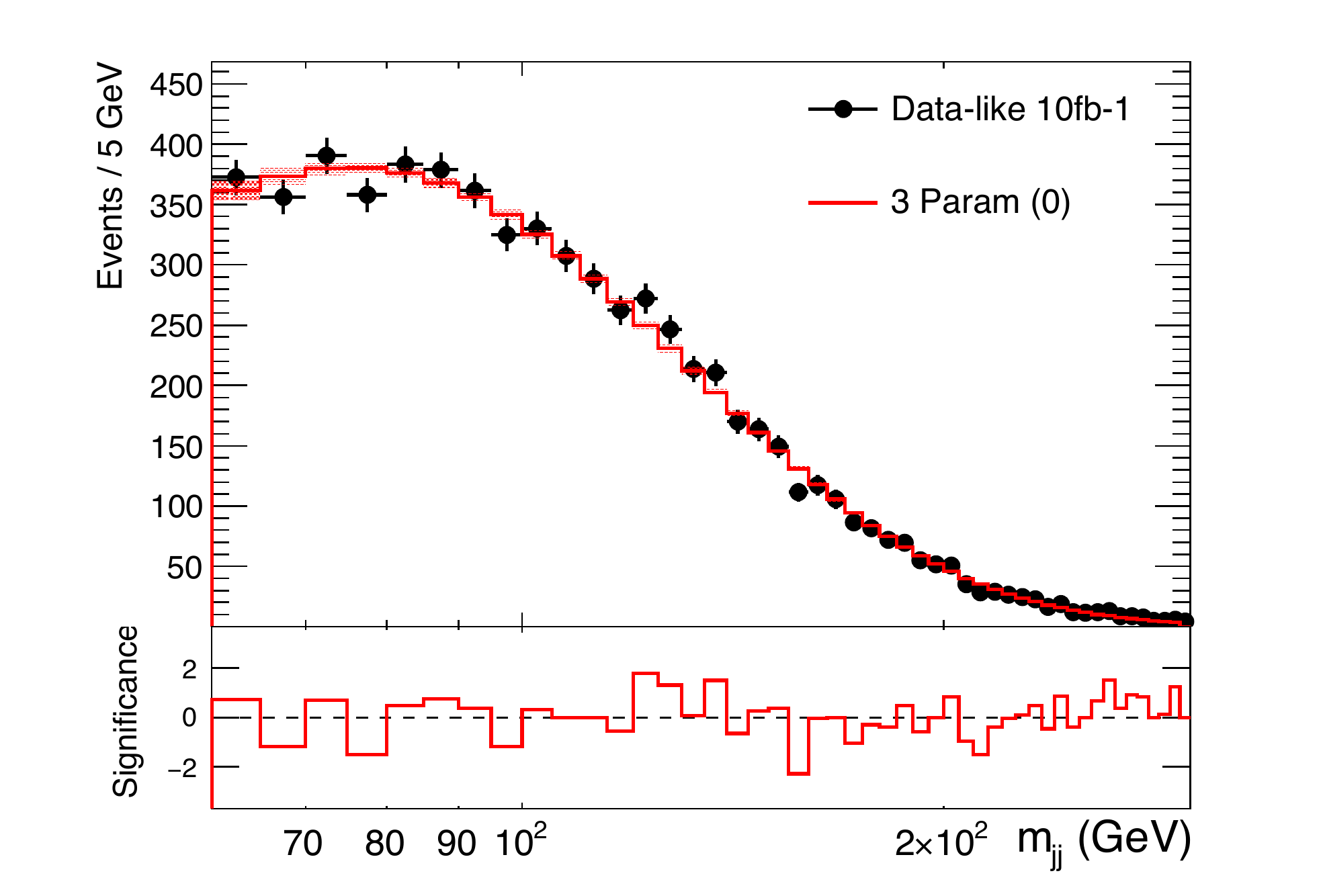}
  \caption{Invariant mass distribution $m_{jj}$ for $Z(\ell^+\ell^-)$+jets at the 14 TeV LHC for (a) MC simulated events normalized to $\rm 10~fb^{-1}$,
and (b) fitted spectrum from three-parameter ansatz function in Eq.~(\ref{eq:fit}) range from 60~GeV to 300~GeV (solid curve).}
  \label{fig:initialtest}
\end{figure}

For the two lepton and $\etmiss$ channel, the dominant background is the SM $Z+jj$ production. With current selection, the two jet invariant mass spectrum is smoothly decreasing within a range of $[60, 300]$~GeV and our signal region lies between 95~GeV and 150~GeV. Making use of the well-measured side-bands, the estimation of background contribution in the signal region could be obtained directly from a fit to the $m_{jj}$ distribution. We generated $Z+$jets samples with MadGraph generator corresponding to $\rm 10\ fb^{-1}$ and passed the events through PYTHIA and DELPHES to simulate the parton shower and ATLAS detector effect. We adopt a parameterization ansatz to fit the distribution in the $m_{jj}$ range from 60~GeV to 300~GeV
\be
f(z)=p_{1}(1-z)^{p_{2}}z^{p_{3}},
\label{eq:fit}
\ee
where $p_{i}$ are free parameters and $z=m_{jj}/\sqrt{s}$. This ansatz is found to provide a satisfactory fit to the generated $Z+$jets MC simulation at 14~TeV, as shown in Fig.~\ref{fig:initialtest}.

\begin{figure}[tb]
  \center
  \includegraphics[width=0.45\linewidth]{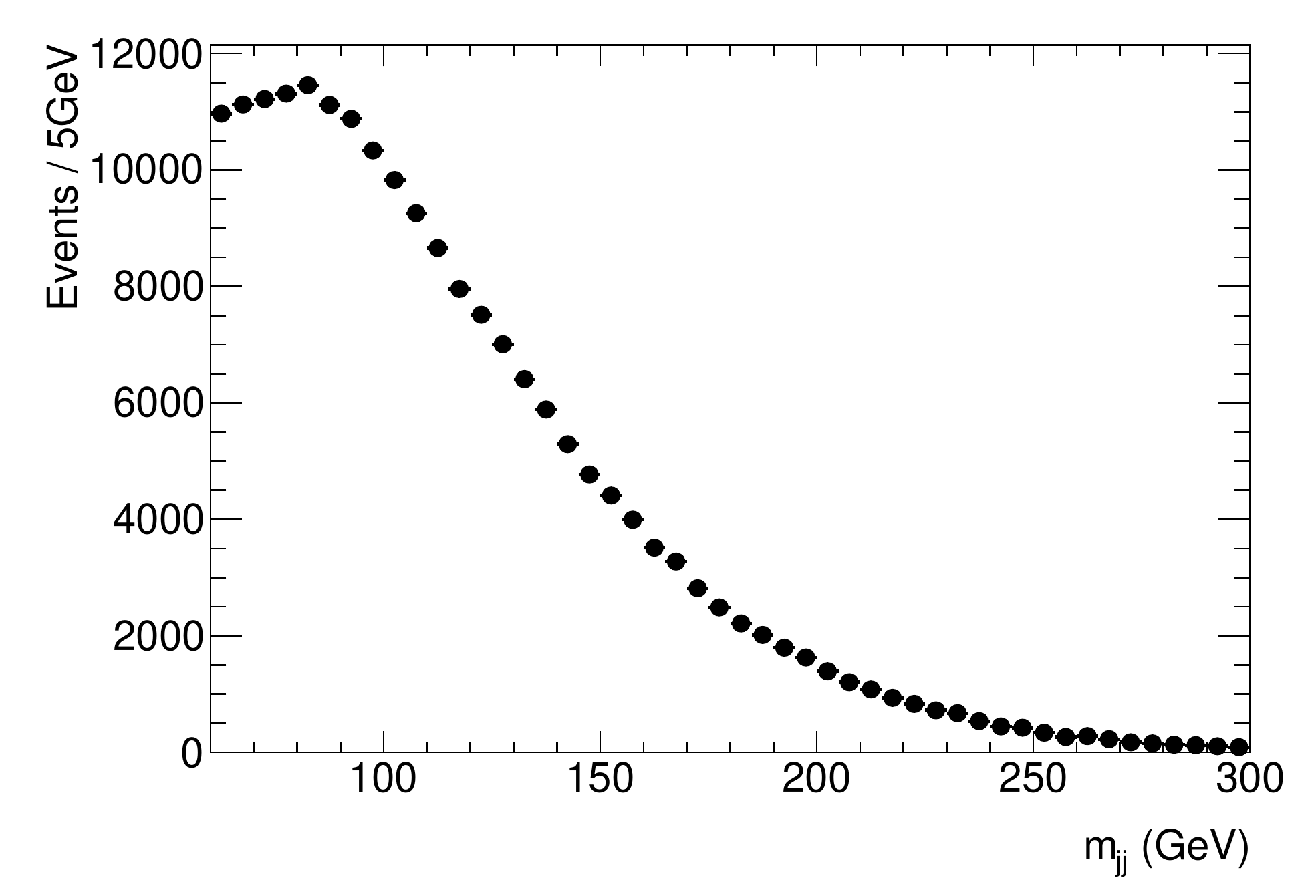}
  \includegraphics[width=0.45\linewidth]{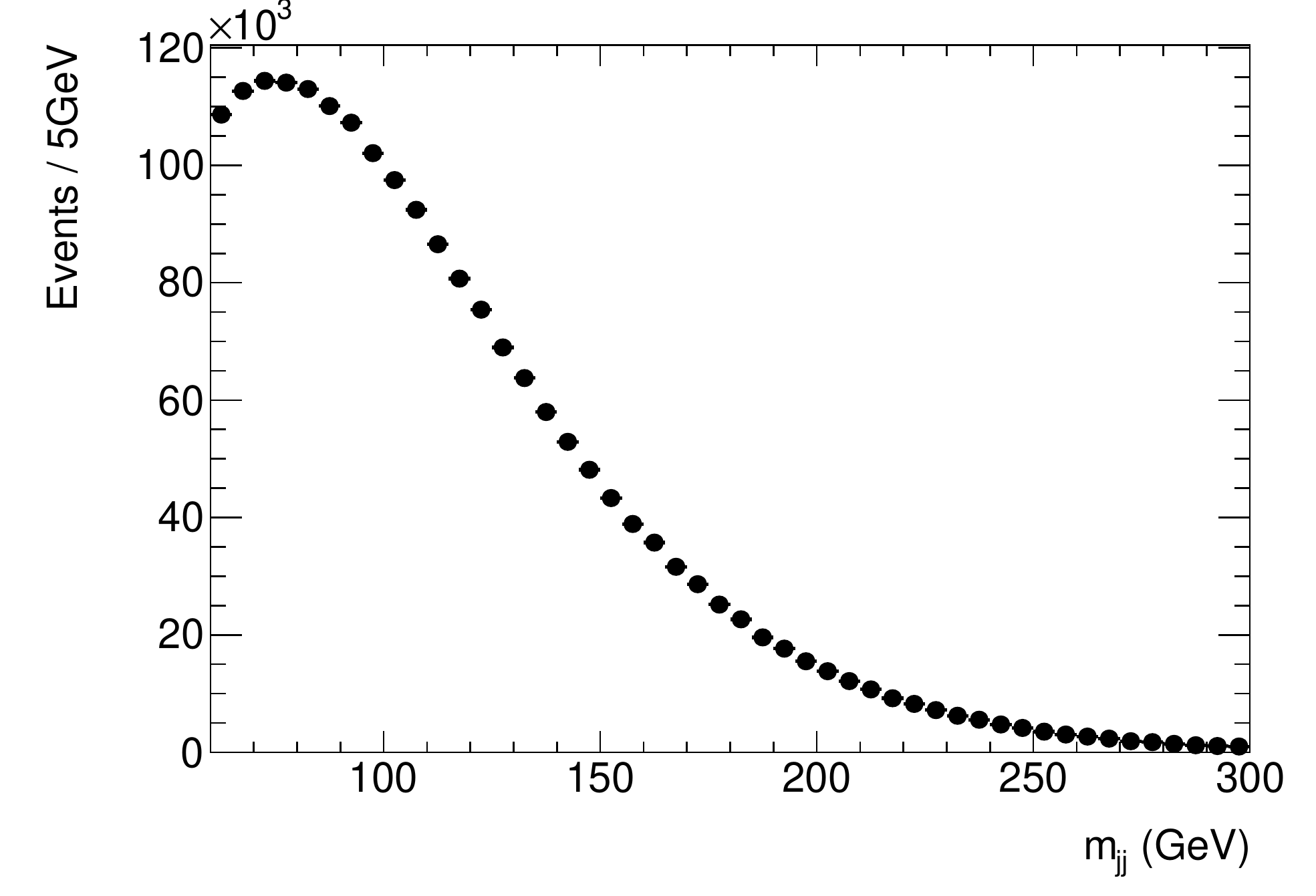}
  \caption{Generated distribution from three-parameter ansatz function  in Eq.~(\ref{eq:fit}) for $m_{jj}$ with (a)
  $\rm 300~fb^{-1}$, (b) and $\rm 3000~fb^{-1}$ (right). }
  \label{fig:spectrum}
\end{figure}

\begin{figure}[tb]
  \center
  \includegraphics[width=0.49\linewidth]{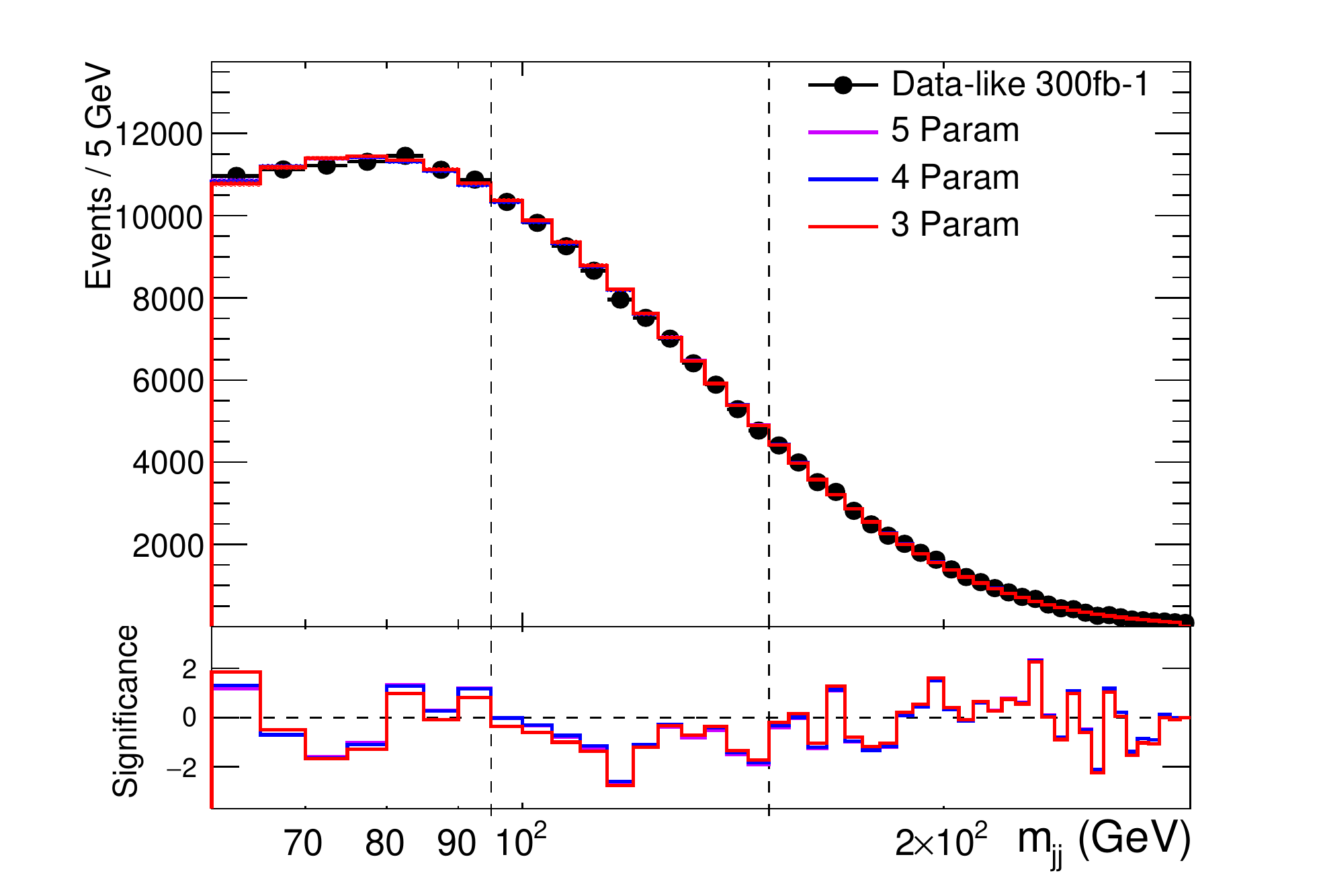}
  \includegraphics[width=0.49\linewidth]{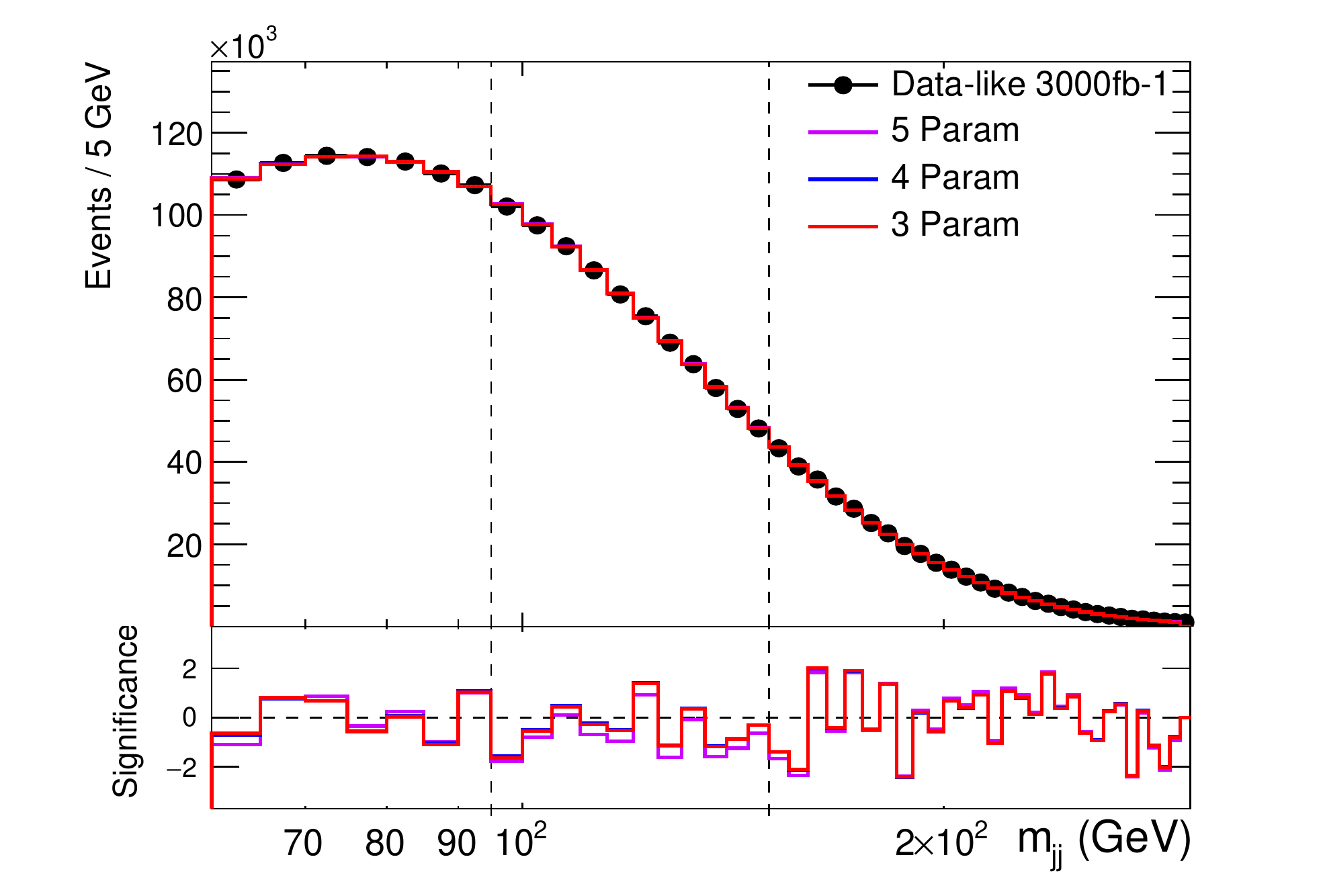}
  \caption{Fitted results for $\rm 300~fb^{-1}$ (left) and $\rm 3000~fb^{-1}$ (right).}
  \label{fig:fitresult}
\end{figure}

In order to estimate the uncertainty of background determination for $\rm 3000~fb^{-1}$ integrated luminosity, we take this three-parameter function  in Eq.~(\ref{eq:fit}) as the baseline to generate the data-like spectrum following Poisson fluctuation. Figure~\ref{fig:spectrum} shows the generated spectra for
$\rm 300~fb^{-1}$ and $\rm 3000~fb^{-1}$.
We fit these spectra with three-parameter, four-parameter and five-parameter functions within the range of $[60, 300]$~GeV but excluding the signal region $[95, 150]$~GeV. The fitting results and uncertainties are summarized in Figure~\ref{fig:fitresult} and Table \ref{tab:fitresult}. Besides the three-parameter function, four-parameter and five-parameter functions are tested as below
\be
f(z)=p_{1}(1-z)^{p_{2}}z^{p_{3}+p_{4}\log(z)},\ \
f(z)=p_{1}(1-z)^{p_{2}}z^{p_{3}+p_{4}\log(z)+p_{5}\log^2(z)}.
\label{eq:fit2}
\ee

\begin{table}
  \center
  \begin{tabular}{l|r|r|r}
    \hline\hline
    Background & $\rm 300~fb^{-1}$ & $\rm 3000~fb^{-1}$ \\\hline
    Expectation & $8.29\times 10^4$ & $8.26\times 10^5$ \\\hline
    3-parameter & $ (8.39 \pm 0.05)\times 10^4 $&$ (8.28 \pm 0.01)\times 10^5$\\
    4-parameter &$ (8.38 \pm 0.05)\times 10^4$&$ (8.27 \pm 0.01)\times 10^5$\\
    5-parameter &$ (8.39 \pm 0.04)\times 10^4$&$ (8.29 \pm 0.01)\times 10^5$\\\hline
    Uncertainty & 1.32\%& 0.21\% \\\hline\hline
  \end{tabular}
  \caption{Fitted results for the background rates from various fitting functions as in Eqs.~(\ref{eq:fit}) and (\ref{eq:fit2}).}
  \label{tab:fitresult}
\end{table}

We also vary the fitting range from $[60, 300]$~GeV to $[70, 250]$~GeV and $[80, 200]$~GeV to test the stability, which are summarized in Table~\ref{tab:fitresult2}. If we consider the variation due to this fitting range as another source of systematics, the uncertainty of background estimation of $Z(\ell\ell)$+jets for $\rm 3000~fb^{-1}$ is 0.33\%.
The uncertainty considered here includes the fitting uncertainty, fitting function variation and fitting range variation, which is largely depending on the statistics of side-band region.
The background uncertainty from fitting is dominated by the statistics of side-band regions, which is proportional to the background yield. To the first-order estimation,
the uncertainties of $Z(\nu\nu)$+jets and $W(\nu\ell)$+jets are comparable at the order of $0.1\%$. We thus summarize the systematic percentage uncertainties for the three leptonic channels as
\be
Z(\ell^+\ell^-)+jj:\ 0.33\% ; \quad
W(\ell^\pm\nu)+jj:\ 0.10\% ; \quad
Z(\nu\nu)+jj:\ 0.13\% .
\label{eq:sys}
\ee

\begin{table}
  \center
  \begin{tabular}{l|c|r|r|r}
    \hline\hline
    $\rm 3000~fb^{-1}$ & True & $[60,300]$~GeV & $[70,250]$~GeV & $[80,200]$~GeV \\\hline
    3-parameter & $8.26\times 10^5$ & $(8.28\pm0.01)\times 10^5$ & $(8.26\pm0.03)\times 10^5$ & $(8.27\pm0.05)\times 10^5$ \\\hline
  \end{tabular}
  \caption{Fitted results for the background rate from various fitting ranges by the fitting function in Eq.~(\ref{eq:fit}).}
  \label{tab:fitresult2}
\end{table}

%

As seen for example in Table~\ref{tab:lvgg-cutflow} for the one-lepton channel, the $t\bar {t}$ background is subdominant yet not negligible. There are other smaller and non-negligible processes such as semi-leptonic decays of di-boson, which are not included in our current studies since they would not change our conclusions.
%
Full simulation and control shall be required on all the relevant processes once the data is available. For our purpose of estimating the signal sensitivity, it suffices to say that the di-jet invariant mass distribution for backgrounds is smooth in the signal region, fitted with simple functions as done above. Since the subdominant backgrounds are statistically much smaller compared to the $Vjj$ process, they would not affect our final results and conclusion.

\section{Alternative Discriminants with Missing Energies}

\label{sec:alternatives}

We note that a momentum balance discriminant has been proposed in Ref.~\cite{Shimmin:2016vlc} as a useful kinematic variable in processes where a new resonant particle is produced in association with a SM vector boson radiated in an initial state, $pp \rightarrow R+V$. The transverse momenta of these states should balance
\begin{equation}
p_T^{R} - p_T^{V} = 0.
\end{equation}
Due to detector effects and radiation, the measured momentum balance is not perfect and it is particularly more severe for the background since the QCD processes tend to have larger radiation. This is a useful kinematic discriminant between the signal and background \cite{Shimmin:2016vlc}. However it is not applicable whenever there is missing energy in the event. In fact, the definition of the missing transverse energy in an event is the negative of the vector sum of the visible $p_T$. In the above example it offers only a tautology for the momentum balance discriminant.  We offer, in the case of events with significant missing energy, a new discriminant to capture the kinematic features of the event.
We define this discriminant by calculating the scalar sum of the transverse momenta of the visible particles in the event, and then subtracting the missing transverse energy
\be
TvQ \equiv \Sigma_i |p_{Ti}| - |\etmiss|.
\ee
This is a version of a momentum balance discriminant, referred as $TvQ$ (Transverse event Quality).  Since the missing momentum in an event is defined by the negative of the vector sum $|\Sigma_i \vec p_{Ti}|$, the quantity $TvQ$ is the difference between the scalar and vector sums of the visible $p_T$ in the event. $TvQ$ tends to be small when the observable particles are a highly collimated collinear bunch, while it takes a large value when the observable particles spread out and when $R+V$ production is near the kinematical threshold.

\begin{figure}[t]
	\centering
	       \includegraphics[scale=0.6]{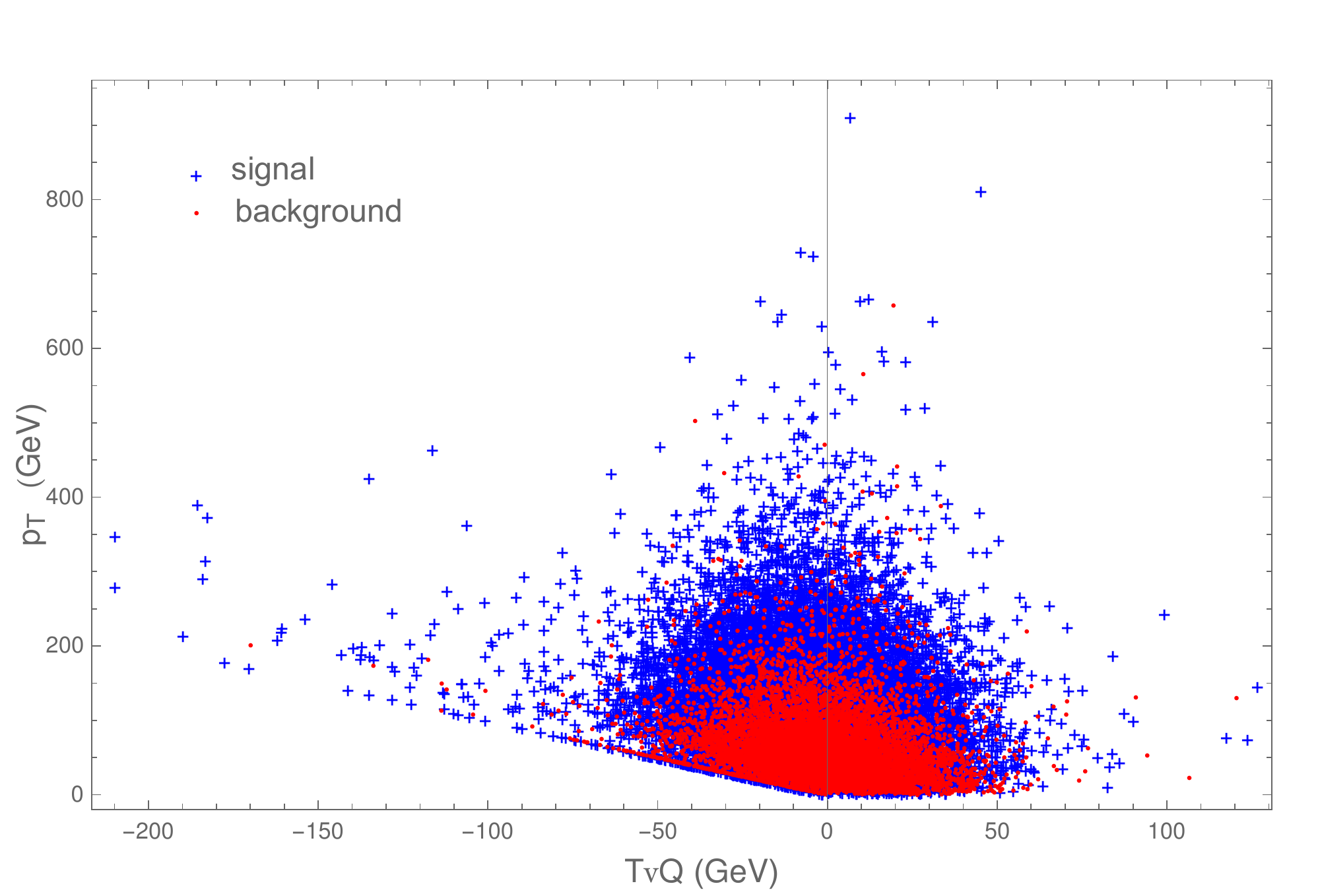}
	\caption[]{Scatter plot of 10000 events for the signal (blue crosses) and background (red dots) in the visible $p_T-TvQ$ plane.
	}
	\label{fig:wd_scatter}
\end{figure}

It would be more intuitive to look at the signal and background in a two dimensional space of discriminants. Consider the $\etmiss$ signal from $pp\to Zh \to \nu\nu\ gg$. We plot the event population in the $p_{T(jj)} - TvQ$ plane as shown in Fig.~\ref{fig:wd_scatter}.  We see that in the signal sample (blue crosses), regions of large visible $p_T$ correlate with the zero value of $TvQ$.  Events with high boost, and therefore columnated Higgs decay products, correlate with lower values of $TvQ$ as predicted. The QCD background sample $Z+$jets (red dots), on the other hand, tends to further spread out.

Another simple discriminant, somewhat correlated with $TvQ$ for the $Zh$ final state is a transverse  angular variable, $\phi_{Zh}$ defined as the angle between the missing transverse energy vector and the vector sum of the visible $p_T$. This is clearly motivated since we expect the $Z$ and $h$ states to be nearly back to back in the event, in contrast to the QCD multiple jet events.
We examined the selective cuts ($-30$ GeV $< TvQ < 10$ GeV) or  ($\pi - 0.5 < \phi_{Zh} < \pi + 0.5$) and found them effective in separating the signal from the backgrounds. In exploiting more kinematical variables in some treatment like Boosted-decision-Tree technique (BDT) or Neural Networks (NN), those discriminative variables may be taken into consideration.

\section{Results and Discussion}
\label{sec:results}

\subsection{Signal significance}
\label{sec:results1}

As we see from the cut-flow tables \ref{tab:lljj-cutflow}-\ref{tab:vvjj-cutflow}, the $Vjj$ backgrounds are dominant.
We calculate the signal statistical significance as
\be
\mathcal{S} = {N_{\rm sig} \over \sqrt{N_{\rm bkg}} },
\ee
with the statistical uncertainty of the dominant background as the only uncertainty.
The combined significance of the $Vh(gg)$ signal is shown in Table \ref{tab:results}.
The three leptonic channels from the $V$ decays give comparable contributions. The two-charged-lepton channel has the smallest signal strength, but cleaner in signal identification. The one and zero-charged-lepton channels show good reconstruction and contribute better sensitivities.
Adding the $0,1,2$ charged-lepton channels, the pure statistical estimation gives a $0.82\sigma$  significance, which indicates how challenging an observation of the SM $Vh(gg)$ signal could be.

When the signal rate and $S/B$ is small, one must worry about the systematic uncertainties for the measurements. As discussed in length in Sec.~\ref{sec:bkg-fit}, we rely on the precision side-band fit to control the systematics in the signal region near $m_{jj}\sim m_h$. If $\epsilon_B$ is the fitted background percentage uncertainty, we then assume the systematic error to be $\epsilon_B\times N_{\rm bkg}$. We thus present a different significance dominated by the systematics, defined as
\be
\mathcal{S}_{\rm sys} = {N_{\rm sig}\over \epsilon_B\times N_{\rm bkg}},
\ee
As shown in Sec.~\ref{sec:bkg-fit}, with 3000 ${\rm fb}^{-1}$ of data and $m_{jj}$ signal mass window taken as $95-150$ GeV, we have $\epsilon_B = 0.33\%,\ 0.10\%,\ 0.13\%$
for the two, one and zero lepton channels, respectively. The results with this significance estimation are also shown in Table \ref{tab:results}. The outcome is worse than the statistical-error-only treatment.
We would also hope the further reduction of non-statistic uncertainties with more dedicated background fitting schemes, once real data is available from experiments.

\begin{table}[tb]
\centering
\begin{tabular}{|c|c|c|c|c|}
\hline
$\sigma$ (fb) & $\ell^+\ell^- + jj$ & $\ell^\pm + \etmiss + jj $ & $\etmiss + jj$ & combined \\ \hline
$Vh$ signal & $7.0\times10^{-2}$ & $4.1\times10^{-1}$& $3.6\times10^{-1}$ & \\ \hline
$Vjj$ background & $2.4\times10^{2}$ & $2.5\times10^{3}$ & $1.6\times10^{3}$ &  \\ \hline
$\mathcal{S}$ & 0.25 & 0.61 & 0.49 & 0.82 \\ \hline
$\mathcal{S}_{\rm sys}$ & 0.09 & 0.17 & 0.17 & 0.26 \\ \hline
\end{tabular}
\caption[]{Signal significance achieved from each channel and combined results for both statistics and systematics dominance.}
\label{tab:results}
\end{table}

\subsection{Bounds on the branching fractions and correlations with $h\to b\bar b,\ c\bar c$}
\label{sec:results2}
The interpretation of these results to bound on individual Higgs decay channels needs further discussion. Thus far, we have only simulated $h\to gg$ as the Higgs decay channel, since it dominates the SM branching fraction of the Higgs decay to light jets. Practically, however, contributions from mis-tagged $h\to b\bar b$, $h\to c\bar c$, and possible light-quark pairs are all accumulated in the events and should be taken into account correlatively. Thus, the signal we have been searching for in this study really is $h\to j'j'$ where $j'$ is an ``un-tagged jet'' including possible $b$, $c$ and $j \ (g,u,d,s)$ contributions.
%

Listed in Table~\ref{tab:eff-bcg} are the working points for the tagging/mis-tagging efficiencies assuming that different observable event categories listed as different rows are un-correlated. For instance, a $b$ quark will be tagged as a $b$ with a probability of $\epsilon_{bb}=70\%$, and mis-tagged as a $c$ and an un-tagged $j'$ with $\epsilon_{cb}=13\%$ and $\epsilon_{j'b}=17\%$, and so on. Here the subscript $a$ denotes the jet-tagged flavor category, and $i$ denotes the parton as the source channel. The numbers are the same as in Category ``$c$-tagging I'' of Table~1
in Ref.~\cite{Perez:2015lra}, as reasonable estimates for the experimental performance at the 14 TeV LHC, and for consistency of later comparison. We extend to the double-tagged event categories with corresponding Higgs branching fraction channels as,
\begin{equation}
e_{ai} = \frac{\epsilon_{ai}^2\times ({\rm BR})_i}{\sum_j\epsilon_{aj}^2\times ({\rm BR})_j}.
\end{equation}
We show in Table~\ref{tab:eff-channel} the percentage contributions of these decay channels $h\to ii$ in each experimentally tagged category $a$. For instance,  a pair of un-tagged jets in category $j'j'$ will have a probability of $74\%$ from the SM Higgs decay to a pair of gluons, and $16\%$ or $10\%$ from $b\bar b$ or $c\bar c$, respectively.
%
With the current tagging efficiency, we translate the significance $0.82\sigma$ on BR$(h\to jj)$ to the un-tagged signal category BR$(h\to j'j')$ by rescaling as 
\be
\mathcal{S}_{j'} ={ \mathcal{S}_j \over e_{j'j} } = { 0.82\sigma \over 74\% } = 1.1\sigma, 
\label{eq:jpjp}
\ee
that accounts for mis-tagged $b\bar b,\ c\bar c$ contributions as well. In other words, if an observation of $h\to j'j'$ were made in the future LHC run, the interpretation for individual channels would be based on  Table~\ref{tab:eff-channel}, with updated tagging efficiencies.

\begin{table}[tb]
    \begin{minipage}{.5\linewidth}
      \caption{Flavor tagging efficiency}
      \label{tab:eff-bcg}
      \centering
 \begin{tabular}{llll}
\hline
$\epsilon_{ai}$ & $b$-quark & $c$-quark & $j=g,u,d,s$ \\
$b$-tag & 70\% & 20\% & 1.25\% \\
$c$-tag & 13\% & 19\% & 0.50\% \\
un-tag $j'j'$ & 17\% & 61\% & 98.25\% \\
\end{tabular}
     \end{minipage}%
    \begin{minipage}{.5\linewidth}
      \centering
      \caption{Fraction of SM decay channels}
      \label{tab:eff-channel}
\begin{tabular}{llll}
\hline
$e_{ai}$ & $h\to b\bar b$ & $h\to c\bar c$ & $h\to jj$ \\
$bb$-tag & 99.6\% & 0.4\% & 0\% \\
$cc$-tag & 90.4\% & 9.6\% & 0\% \\
un-tag $j'$ & 16\% & 10\% & 74\% \\

\end{tabular}
    \end{minipage}
\end{table}

\begin{figure}[tb]
  \center
  \includegraphics[width=0.32\linewidth]{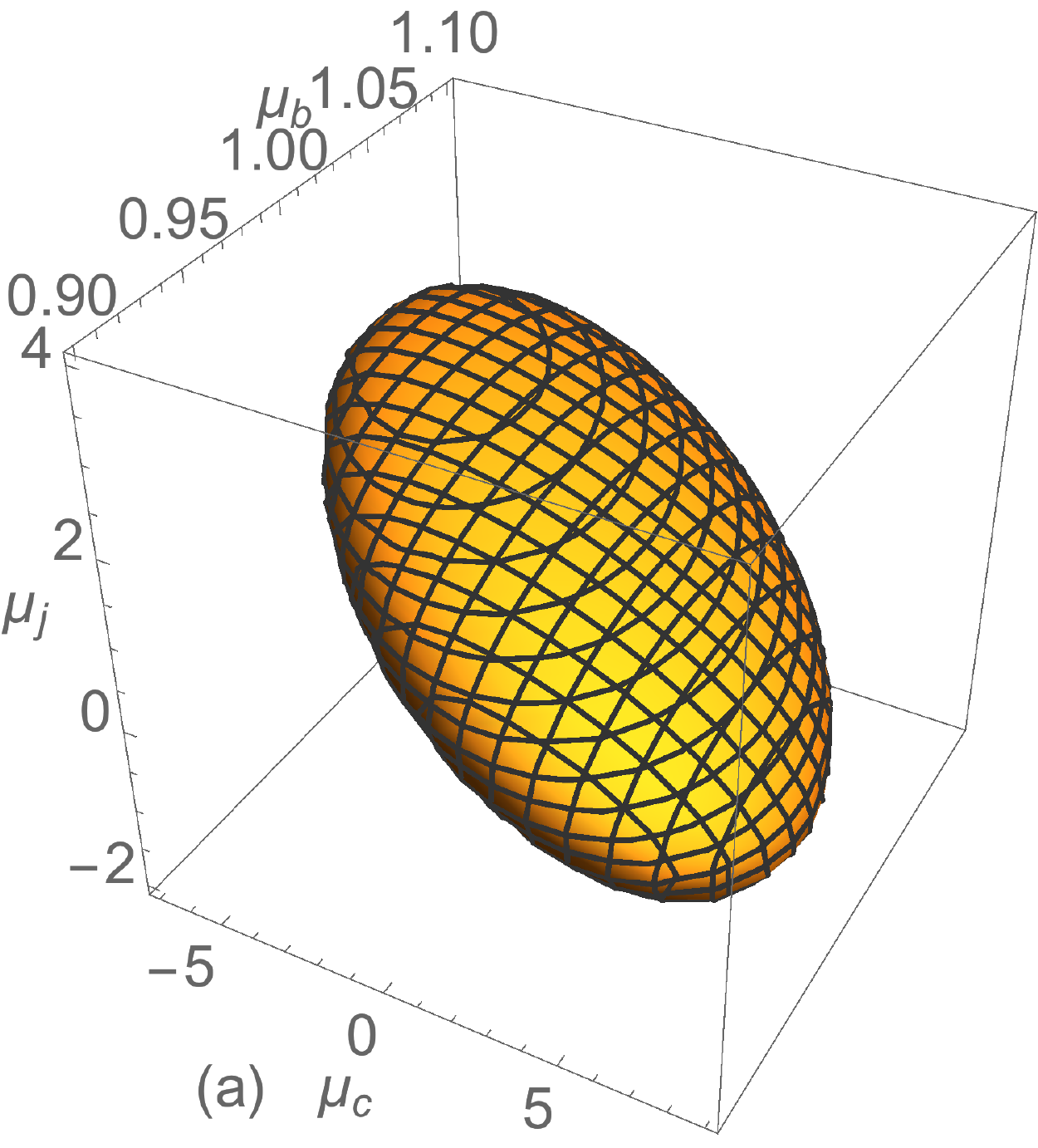}
  \includegraphics[width=0.33\linewidth]{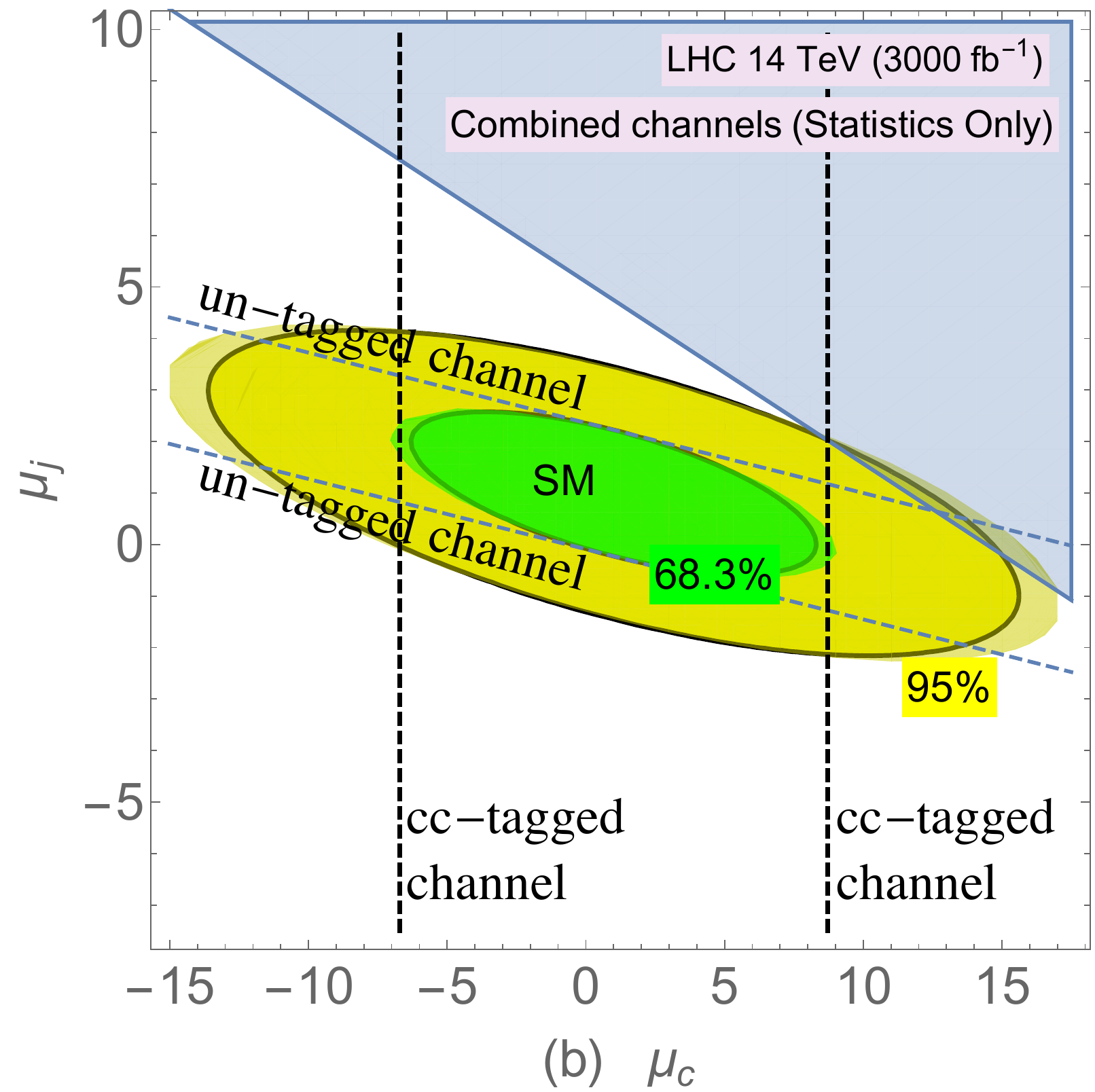}
  \includegraphics[width=0.33\linewidth]{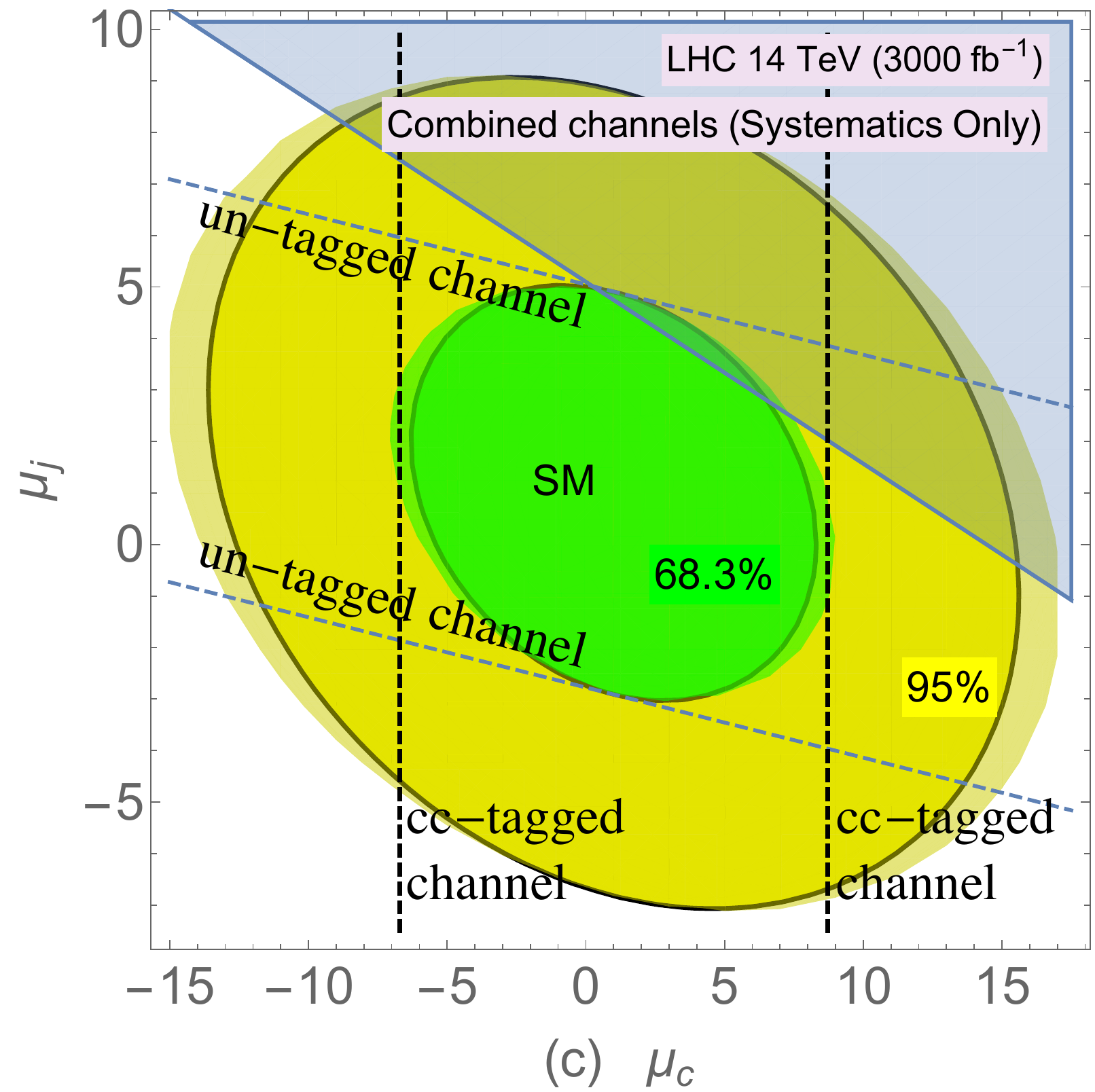}
  \caption{Signal strengths in correlated regions for (a) $1\sigma$ contour in 3-dimension ($\mu_b$, $\mu_c$, $\mu_j$), (b) and (c) contours in $\mu_c$-$\mu_j$ plane, for statistics only and including systematic uncertainties, respectively. The shadowed contour regions are the projection of the 3D contour ($\mu_b$, $\mu_c$, $\mu_j$) onto the $\mu_c$-$\mu_j$ plane at $1\sigma$ and $2\sigma$, and the solid ovals are for fixing $\mu_b =1$. The grey triangle area at the upper right corner is unphysical BR($h\to bb$) + BR($h\to cc$) + BR($h\to jj$) $> 1$.}
  \label{fig:ucug-contour}
\end{figure}

As is customary, we define the signal strength for a decay channel $h\to ii$ as
\be
\mu_i = \frac{{\rm BR}(h\to i i)}{{\rm BR}^{\rm SM}(h\to i i)},
\ee
where we consider $ii=b\bar b,\ c\bar c,$ and $jj$.
Assuming each category is statistically independent and following Gaussian statistics. We combine the three categories to get the three dimensional contour constraint on $\{\mu_b, \mu_c, \mu_j\}$ correlatively based on the relation
\begin{equation}
\begin{aligned}
\mathcal{S}^2 &> \sum_a \chi_a^2 = \sum \frac{(x_a-\overline{x}_a)^2}{\sigma^2_a}\\
&= \sum_a {(\sum_{i} \epsilon_{ai}^2 {\rm BR}_i N_{\rm sig}^{prod} - \sum_i \epsilon_{ai}^2 {\rm BR}_i^{\rm SM} N_{\rm sig}^{prod})^2 \over (\sqrt{N_{\rm bkg}})^2} \\
&= \sum_a {(\sum_i e_{ai}\ \mu_i - 1)^2 \over (1/\mathcal{S}_a)^2}
\end{aligned}
\end{equation}
where $\mathcal{S}_a$ is the significance from each category identified by experiments, and $e_{ai}$ are the double efficiencies from each decay channel $i$ in category $a$ given in Table \ref{tab:eff-channel}.\footnote{Note the different efficiencies defined in Tables \ref{tab:eff-bcg} and \ref{tab:eff-channel}, with the normalizations $\sum_a \epsilon_{ai}=1$ in categories, and $\sum_i e_{ai}=1$ in channels.} We take $\mathcal{S}_a$ =
($11,\ 1.35,\ 1.1\ (0.35)$) for the three categories, assuming only statistical errors with $3000~ {\rm fb}^{-1}$ data. The first number is from Table 12 in the ATLAS MC study \cite{atlasstudy}, making use of ``One+Two-lepton" combined sensitivity. The second number comes from Fig.~2(a) of Ref.~\cite{Perez:2015lra}, the extrapolated study on the same MC dataset assuming the same tagging efficiency. Assuming most of the sensitivity on $\mu_c$ comes from the double $c$-tagged category, we likewise rescale the number with $e_{c'c}$ and a $\sqrt{2}$ since they consider $2\times 3000\ \rm fb^{-1}$ data from two experiments. The third number is from our current ``Zero+One+Two-lepton" un-tagged jets study, with the number in parenthesis including the systematic error. The fully correlated signal strengths are plotted in Fig.~\ref{fig:ucug-contour}, for (a) a 3-dimensional contour in ($\mu_b$, $\mu_c$, $\mu_j$) at $1\sigma$, (b) the projected  contour on the  $\mu_j-\mu_c$ plane with statistical error only, and (c) with systematical error dominance. 
The shadowed contour regions are the projection of the 3D contour ($\mu_b$, $\mu_c$, $\mu_j$) onto the $\mu_c$-$\mu_j$ plane at $1\sigma$ and $2\sigma$, and the solid ovals are for a fixed value $\mu_b =1$. Allowing $\mu_b$ to float, the contour regions are slightly larger than the ovals. 
We note that certain values of the parameter space plane are excluded when BR($h\to bb$) + BR($h\to cc$) + BR($h\to jj$) $> 1$ and where our SM production assumption breaks down. This is represented in the plots by the gray shaded region. The 95\% Confidence Level (CL) global upper bounds (approximately 2$\sigma$) on the branching fractions with statistical errors (systematic errors) for 3000 fb$^{-1}$ with respect to the SM predictions can be obtained as
\bea
\label{eq:jj}
&& {\rm BR}(h\to jj) \leq 4~(9)\times {\rm BR}^{SM}(h\to gg) , \\
&& {\rm BR}(h\to c\bar c) < 15\times {\rm BR}^{SM}(h\to c\bar c),
\label{eq:cc}
\eea
Although this bound on the $h\to gg$ channel is not nearly as strong as that from the production fit $gg\to h$ assuming the SM value, our study and results lay out the attempt of the search for the direct decay of the Higgs boson to gluons and the light quarks. The result for $c\bar c$ is comparable with the best existing extrapolations \cite{Bishara:2016aa, Perez:2015lra}, although adding the un-tagged category slightly improve the constraints on the $c$-quark Yukawa coupling, as expected.

%

Further improvements can be made by including the production of the vector boson fusion 
(VBF) \cite{Collaboration:2015aa} and $t\bar t h$ \cite{Collaboration:2016aa}. They are the sub-leading contributions to the $h\to jj$ study at Run I and become more important production channels at Run II \cite{CMS:2016mmc}. Our study includes for simplicity only double-tagged categories, and single $b$ or $c$ tagged categories can be further included as done in the recast by Ref.~\cite{Perez:2015aa}. Statistics can be further improved by analysis with likelihood fitting, BDT, etc. once data is available.

\subsection{Bounds on light-quark Yukawa couplings}
\label{sec:results3}

So far, possible contributions from light quarks ($u,d,s$) have been ignored in accordance with the SM expectation. The bound on $h\to jj$ in Eq.~(\ref{eq:jj}) can be translated into those for the light quark Yukawa couplings. Assuming the SM $ggh$ coupling, and varying one light quark Yukawa $y_q$ at a time, we translate our bound on $\mu_j$ to the Yukawa couplings for light quarks ${u, d, s}$  by scaling the branching fraction with $\mu_q \propto y_q^2$. Our results of the bounds on the Yukawa couplings normalized to $y_b$ are shown in Table~\ref{tab:compare}.
There have been attempts to probe the light quark Yukawa couplings in the literature \cite{Zhou:2015wra, Bishara:2016aa,Soreq:2016aa, Bonner:2016sdg}.
Recent studies on the inclusive Higgs production and its spectra of $p_{T(h)}$ and $y_h$ claim various improved constraints on the couplings \cite{Bishara:2016aa,Soreq:2016aa}, compared to constraints from a global fit \cite{Kagan:2014ila}.
The upper bounds from our study of Higgs decay to light jets are comparable to those derived from the Higgs production kinematics, as also shown in Table~\ref{tab:compare}, and thus provide complementary information to the existing approaches. We also see from the table that our result may offer a better probe to the strange-quark Yukawa coupling.

\begin{table}[tb]
\centering
\begin{tabular}{|c|c|c|c|}
 \hline
$\mathcal{L} (\rm fb^{-1})$ & $\overline{\kappa}_u$ & $\overline{\kappa}_d$ & $\overline{\kappa}_s$  \\ \hline
300 (un-tagged $j'j'$) & 1.3 & 1.3 & 1.3\\ \hline
3000 (un-tagged $j'j'$) & 0.6 & 0.6 & 0.6 \\ \hline
\hline
Current Global Fits \cite{Kagan:2014ila} & 0.98  & 0.97& 0.70 \\ \hline
300 \cite{Soreq:2016aa}& 0.36  & 0.41 & \\ \hline
3000 \cite{Bishara:2016aa} &  &  & 1  \\ \hline
\end{tabular}
\caption[]{Extrapolated upper bounds at $95\%$ CL on the light-quark Yukawa couplings $\overline{\kappa}_q = {y_q}/{y_b^{\rm SM}} (q=u,d,s)$}
\label{tab:compare}
\end{table}
%

%


\section{Summary and Conclusions}
\label{sec:conclude}

We have carried out a detailed study of the Higgs boson decay to light un-tagged jets in the vector boson associated channel $pp \rightarrow  V h$, with $h\to gg$ and $V=W^\pm,\ Z$ decaying to leptons at the 14 TeV HL-LHC with 3000 fb$^{-1}$.  To differentiate the di-jet signal from the huge SM QCD backgrounds, we have maximized the signal sensitivity by combining searches in the 0, 1 and 2-leptonic decay channels of the vector bosons. We used MadGraph, PYTHIA, and DELPHES for the signal and background simulations. Our findings can be summarized as follows.
\begin{itemize}
\item
In Sections \ref{sec:2lepton}-\ref{sec:0lepton}, we optimized the kinematical cuts according to the individual signal channels to enhance the $S/\sqrt B$ as well as $S/B$. The boosted kinematics for the di-jet signal has the advantage to improve $S/B$, while to keep the $S/\sqrt B$ roughly the same.
We proposed the ``di-jet-vicinity" Higgs mass reconstruction method as seen in Fig.~\ref{fig:hmass-reconst}, and tested its effectiveness against the pile-up effects as in Fig.~\ref{fig:pile-up}.
\item
In Sec.~\ref{sec:bkg-fit}, we studied in great detail on how to control the systematic errors by making use of the side-bands with a few fitting functions. We found that with 3000 fb$^{-1}$, it is conceivable to  achieve the sub-percent level systematic uncertainties, as given in Eq.~(\ref{eq:sys}). It would be crucially important to take advantage of the large statistics and to keep the systematics under control.
\item
We may reach about $1\sigma$ combined significance for the un-tagged di-jet channel, as shown in Table~\ref{tab:results} and in Eq.~(\ref{eq:jpjp}). We also considered the correlation with mis-tagged events from $h\to b\bar b,\ c\bar c$ channels, as discussed in Sec.~\ref{sec:results2}
\item
Assuming the SM $Vh$ production,
our results can be translated to upper bounds on the branching fractions
of $4$ and $15$ times the SM values for BR($h\to gg$) and BR($h\to c\bar c$), respectively,  at 95\% CL, seen in Eqs.~(\ref{eq:jj}) and (\ref{eq:cc}).
\item
Exploiting our results, indirect upper bounds on light-quark Yukawa couplings can be extracted, as summarized in Table \ref{tab:compare}, and compared with the currently existing literature.
\item
We pointed out that there are other variables to explore. Kinematic
discriminants like $TvQ$ and $\phi_{Zh}$ as discussed in Sec.~\ref{sec:alternatives} may be among them.
In the hope to improve the simple cut-based analyses, multiple variable methods like BDT and NN would be promising. Addition of other production channels such as VBF and $t\bar t h$ will also help to strengthen the bounds.
\end{itemize}

After the Higgs boson discovery and initial measurements for the SM-like properties at the LHC Run I and Run II, it is imperative at the HL-LHC to tackle the more challenging channels with the rare Higgs decays. Our studies on the Higgs decay to the light un-tagged jets would hopefully serve as an initial proposal among the future efforts.

\acknowledgments{
The work of T.H.~and Z.Q.~was supported in part by the U.S.~Department of Energy under grant No.~DE-FG02-95ER40896, in part by PITT PACC. Z.Q.~was also supported in part by a  PITT PACC Predoctoral Fellowship from School of Art and Science at University of Pittsburgh.
The work of K.H.~was supported by the National Science Foundation Graduate Research Fellowship under Grant No.~DGE-134012.
The work of L.C.~and K.H.~was supported by the U.S.~Department of Energy under grant No.~DE-SC0013529.}

\bibliographystyle{jhep}
\bibliography{references}

\end{document}